\pdfoutput=1
\documentclass[english]{IEEEtran}
\usepackage[LGR,T1]{fontenc}
\usepackage[latin9]{inputenc}
\usepackage{xcolor}
\usepackage{array}
\usepackage{float}
\usepackage{booktabs}
\usepackage{bm}
\usepackage{multirow}
\usepackage{amsmath}
\usepackage{amsthm}
\usepackage{amssymb}
\usepackage{graphicx}
\PassOptionsToPackage{normalem}{ulem}
\usepackage{ulem}

\makeatletter

\DeclareRobustCommand{\greektext}{%
  \fontencoding{LGR}\selectfont\def\encodingdefault{LGR}}
\DeclareRobustCommand{\textgreek}[1]{\leavevmode{\greektext #1}}

\newcommand{\lyxmathsym}[1]{\ifmmode\begingroup\def\b@ld{bold}
  \text{\ifx\math@version\b@ld\bfseries\fi#1}\endgroup\else#1\fi}

\providecommand{\tabularnewline}{\\}
\floatstyle{ruled}
\newfloat{algorithm}{tbp}{loa}
\providecommand{\algorithmname}{Algorithm}
\floatname{algorithm}{\protect\algorithmname}
\providecolor{lyxadded}{rgb}{0,0,1}
\providecolor{lyxdeleted}{rgb}{1,0,0}
\DeclareRobustCommand{\mklyxadded}[1]{\textcolor{lyxadded}\bgroup#1\egroup}
\DeclareRobustCommand{\mklyxdeleted}[1]{\textcolor{lyxdeleted}\bgroup\mklyxsout{#1}\egroup}
\DeclareRobustCommand{\mklyxsout}[1]{\ifx\\#1\else\sout{#1}\fi}
\theoremstyle{remark}
\newtheorem*{notation*}{\protect\notationname}
\theoremstyle{plain}
\newtheorem{thm}{\protect\theoremname}
\theoremstyle{plain}
\newtheorem{assumption}{\protect\assumptionname}
\theoremstyle{plain}
\newtheorem{lem}{\protect\lemmaname}
\theoremstyle{plain}
\newtheorem{cor}{\protect\corollaryname}
\theoremstyle{definition}
\newtheorem{defn}{\protect\definitionname}

\usepackage{amsmath}
\usepackage{amssymb}
\usepackage{algorithm}
\usepackage{algpseudocode}

\usepackage[level=3]{wgroup_message}  % set level to 0, to hide all notes
\usepackage{graphicx,psfrag,cite,subfigure}
\author{
\IEEEauthorblockN{}

\IEEEauthorblockA{Jijia Tian, Wangqian Chen, Junting Chen, Pooi-Yuen Kam\\
School of Science and Engineering and Future Network Intelligence Institute (FNii)\\
The Chinese University of Hong Kong, Shenzhen, Guangdong 518172, China}
}


\usepackage[acronym]{glossaries}
\newcommand{\newac}{\newacronym}
\newcommand{\ac}{\gls}

\newcommand{\acpl}{\glspl}

\usepackage{placeins}
\makeglossaries

\newac{speb}{SPEB}{square position error bound}
\newac[plural=EFIMs,firstplural=Fisher information matrices (EFIMs)]{efim}{EFIM}{Fisher information matrix}
\newac{ne}{NE}{Nash equilibrium}
\newac{mse}{MSE}{mean squared error}
\newac{toa}{TOA}{time-of-arrival}
\newac{snr}{SNR}{signal-to-noise ratio}
\newac{lan}{LAN}{local area network}
\newac{psd}{PSD}{positive semidefinite}
\newac{pd}{PD}{positive definite}
\newac{wrt}{w.r.t.}{with respect to}
\newac{lhs}{L.H.S.}{left hand side}
\newac{wp1}{w.p.1}{with probability 1}
\newac{kkt}{KKT}{Karush-Kuhn-Tucker}
\newac{wlog}{w.l.o.g.}{without loss of generality}
\newac{mle}{MLE}{maximum likelihood estimation}
\newac{gps}{GPS}{global positioning system}
\newac{rssi}{RSSI}{received signal strength indicator}
\newac{mimo}{MIMO}{multiple-input multiple-output}
\newac{csi}{CSI}{channel state information}
\newac{fdd}{FDD}{frequency division duplexing}
\newac{ms}{MS}{mobile station}
\newac{bs}{BS}{base station}
\newac{d2d}{D2D}{device-to-device}
\newac{slnr}{SLNR}{signal-to-interference-leakage-and-noise-ratio}
\newac{ula}{ULA}{uniform linear antenna array}
\newac{pas}{PAS}{power angular spectrum}
\newac{mmse}{MMSE}{minimum mean square error}
\newac{zf}{ZF}{zero-forcing}
\newac{rzf}{RZF}{regularized zero-forcing}
\newac{as}{AS}{angular spread}
\newac{aod}{AOD}{angle of departure}
\newac{iid}{i.i.d.}{independent and identically distributed} 
\newac{sinr}{SINR}{signal-to-interference-and-noise ratio}
\newac{tdd}{TDD}{time-division duplex}
\newac{rvq}{RVQ}{random vector quantization}
\newac{rhs}{R.H.S.}{right hand side}
\newac{mrc}{MRC}{maximum ratio combining}
\newac{cdf}{CDF}{cumulative distribution function}
\newac{a.s.}{a.s.}{almost surely}
\newac{los}{LOS}{line-of-sight}
\newac{jsdm}{JSDM}{joint spatial division and multiplexing}
\newac{map}{MAP}{maximum a posteriori}
\newac{klt}{KLT}{Karhunen-Lo\`eve Transform}
\newac{lbe}{LBE}{link bargaining equilibrium}
\newac{se}{SE}{Stackelberg equilibrium}
\newac{uav}{UAV}{unmanned aerial vehicle}
\newac{nlos}{NLOS}{non-line-of-sight}
\newac{pdf}{PDF}{probability density function}
\newac{em}{EM}{expectation-maximization}
\newac{knn}{KNN}{$k$-nearest neighbor}
\newac{svd}{SVD}{singular value decomposition}
\newac{nmf}{NMF}{non-negative matrix factorization}
\newac{umf}{UMF}{unimodality-constrained matrix factorization}
\newac{rmse}{RMSE}{rooted mean squared error}
\newac{olos}{OLOS}{obstructed line-of-sight}
\newac{mmw}{mmW}{millimeter wave}
\newac{ber}{BER}{bit error rate}
\newac{rss}{RSS}{received signal strength}
\newac{lp}{LP}{linear program}
\newac{ufw}{U-FW}{unimodal Frank-Wolfe}
\newac{utf}{UTF}{unimodality-constrained tensor factorization}
\newac{fw}{FW}{Frank-Wolfe}
\newac{iot}{IoT}{Internet-of-Things}
\newac{mae}{MAE}{mean absolute error}
\newac{crb}{CRB}{Cram\'er-Rao bound}
\newac{aoa}{AoA}{angle of arrival}
\newac{wcl}{WCL}{Weighted Centroid Localization}
\newac{fel}{FL}{Federated Learning}
\newac{ms2}{MS}{Mean Shift}
\newac{dp}{DP}{Differential Privacy}
\newac{adp}{ADP}{Adaptive Differential Privacy}
\newac{g-adpfl}{G-ADPFL}{Geometric-Adaptive Differential Privacy Federated Learning}
\newac{fedavg}{FedAvg}{Federated Averaging}
\newac{dpfl}{DPFL}{Differential Privacy Federated Learning}

\setkeys{Gin}{width=1.0\columnwidth}

\makeatother

\usepackage{babel}
\providecommand{\assumptionname}{Assumption}
\providecommand{\corollaryname}{Corollary}
\providecommand{\definitionname}{Definition}
\providecommand{\lemmaname}{Lemma}
\providecommand{\notationname}{Notation}
\providecommand{\theoremname}{Theorem}

\begin{document}
\title{Geometry-Aligned Differential Privacy for Location-Safe Federated
Radio Map Construction}

\maketitle
%% The following is the formatting requirement of TWC submission
%
% 12pt, draftclsnofoot, peerreview, a4paper, oneside, twocolumn
%
% ================
%% The following commnds automatically adjust the size of the figures and the font size on the figures according to single/double column confirguration. However, it does not affect the figures with their size exiplicitly specified. It also requrie \figfontsize command to be put in the psfrag block.
%% Choice of font size: tiny, scriptsize, footnotesize, small, normalsize, large, Large, LARGE, huge Huge

%% Uncomment the following if a single column format is used.

%\newcommand*{\SINGLECOLUMN}{}

\ifdefined\SINGLECOLUMN
	% - single column setting
	\setkeys{Gin}{width=0.5\columnwidth}
	\newcommand{\figfontsize}{\footnotesize} 
\else
	% - double column setting
	\setkeys{Gin}{width=1.0\columnwidth}
	\newcommand{\figfontsize}{\normalsize} 
\fi
% ================
%% The following define annotation formate
% TBD

\begin{abstract}
Radio maps that describe spatial variations in wireless signal strength
are widely used to optimize networks and support aerial platforms.
Their construction requires location-labeled signal measurements from
distributed users, raising fundamental concerns about location privacy.
Even when raw data are kept local, the shared model updates can reveal
user locations through their spatial structure, while naive noise
injection either fails to hide this leakage or degrades model accuracy.
This work analyzes how location leakage arises from gradients in a
virtual-environment radio map model and proposes a geometry-aligned
\ac{dp} mechanism with heterogeneous noise tailored to both confuse
localization and cover gradient spatial patterns. The approach is
theoretically supported with a convergence guarantee linking privacy
strength to learning accuracy. Numerical experiments show the approach
increases attacker localization error from $30$ m to over $180$
m, with only $0.2$ dB increase in radio map construction error compared
to a uniform-noise baseline.
\end{abstract}

\section{Introduction}

Radio maps are models and structured data that capture radio characteristics
in a specific geographic environment, and they can play an important
role in optimizing the performance of communication networks\cite{ZenCheXuWu:J24,9994774,chen2024diffraction,10041012}.
In \ac{mimo} communication systems, one may exploit radio maps to
organize historical data to assist channel modeling, enhanced \ac{mimo}
beam tracking, and efficient interference management\cite{10287775,10108969,9634111,10486853}.
Furthermore, in low-altitude \ac{uav} networks, radio maps provide
comprehensive spatial data that enables flight path planning, reliable
communication under possible blockage, and joint communication, sensing
and safe flight control\cite{ZenXuJinZha:J21,10653723,ZhaZha:J21}.

However, it is challenging to construct up-to-date radio maps, because
the environment that affects the signal propagation may be slowly
varying in time. As a result, it is essential to keep collecting fresh
radio measurement data to monitor the propagation environment\cite{10335642,9523765,9500705,10445518,9992123}.
Consequently, a practical remedy is to exploit the {\em crowd-sourcing}
mechanism that motivates a massive number of mobile terminals to measure
and report the channel data they observe to the fusion center. However,
a conventional crowd-sourcing radio surveying would impose a critical
concern on location privacy, where the mobile terminals are required
to report their locations as labels of the reported channel measurement
data.

\emph{Is it possible to hide the location information in data collection
for radio map construction?} \ac{fel} provides a generic mechanism
to train a model distributively and collaboratively without uploading
the raw data\cite{10771980,9855842,10255285,10669237,10615418,10460129}.
Specifically, the global radio map model is maintained by a central
server, and a subset of model parameters are shared among mobile users;
the users update the parameters of the local model based on up-to-date
local measurement data, and transmit the updated parameters, instead
of the raw data, to the central server for model aggregation\cite{mcmahan2017communication,pmlr-v97-yurochkin19a}.

However, as radio map models usually contain geographically structured
parameter sets\cite{10437033,10041012} or are themselves geographically
structured data arrays\cite{9992123,10335642}, uploading the model
parameters may still expose the location information of the users,
although the original measurement data is kept locally. For example,
in a joint radio map and environment map construction problem\cite{chen2024diffraction,10041012},
the model parameters consist of the heights of the obstacles in the
neighborhood of the users, and therefore, updating these environment
parameters can reveal the location information of the users. As a
result, both the original measurement data and the intermediate parameters
need to be protected.

While conventional \ac{fel} may pose a potential risk of leaking
location information, a \ac{dp} strategy may be considered, where
zero-mean noise is added to the parameters uploaded to the central
server. As a result, the location information hidden in the modeled
parameters is scrambled; in addition, when there are a large number
of users participating in the federated radio map construction, the
zero-mean noise in the update from these users can be averaged out
at the central server. However, it is still not clear how to control
the noise for both protecting the location privacy and not significantly
reducing the learning performance of the radio map under \ac{fel}.

In this paper, we study the issue of protecting the location privacy
of the users in collaborative radio map construction, where the users
collect radio measurement data locally under a crowd-sourcing mechanism,
and collaboratively train a radio map model under the coordination
of a central server. This paper is an extended version of \cite{10615418}.\footnote{This work was presented in part at the IEEE International Conference
on Communications Workshops (ICC Workshops), Denver, CO, USA, 2024.} We first demonstrate and analyze the location information leakage
in the classical \ac{fel} scheme. Then, we adopt a geometry-aligned
\ac{dp} scheme to protect the location information by adding zero-mean
noise to the parameters uploaded to the central server. Specifically,
we develop a noise-adding strategy to protect location information
from {\em two} perspectives. First, the spatial distribution of the
differential privacy noise is deliberately chosen to follow the geometric
structure of the gradient field, rather than being uniformly random,
such that the localization error of the adversary is maximized. Second,
the noise parameters are also tuned to hide the spatial pattern of
the noise corrupted gradient field to further protect the location
information. A regularized objective is developed to achieve the two
competing goals in the proposed geometry-aligned \ac{dp} framework.
We derive the convergence proof for the \ac{fel} algorithm under
such \ac{dp} scheme and reveal the relation between the convergence
and the noise upper bound.

To summarize, this paper makes the following contributions:
\begin{itemize}
\item We analyze how the location information may leak from the local gradient
for a joint radio map and environment sensing problem. Mathematically,
we show that the local gradient tends to decrease in magnitude for
the environment-related parameters that are spatially distant from
the user. This property implies a simple adversary localization strategy
from the local gradients.
\item We propose a geometry-aligned \ac{dp} framework that applies heterogeneous
noise to individual parameters in the local gradients. The noise levels
are dynamically optimized to maximize the localization error of potential
adversarial inference attacks, while simultaneously minimizing the
spatial variance of the gradient field. The dual objective safeguards
the location-specific spatial patterns embedded in the gradients,
while preserving their utility for learning.
\item We derive a convergence bound for the proposed federated radio map
construction framework with geometry-aligned \ac{dp} protection.
The bound quantifies the trade-off between the learning performance
of the radio map model and the strength of location privacy protection.
\item Numerical experiments on synthetic urban radio map datasets show that
the proposed geometry-aligned \ac{dp} scheme increases the adversary
localization \ac{rmse} from about $30$m to over $180$m, while degrading
the radio map reconstruction \ac{mae} by only about $0.2$dB compared
with a conventional uniform-noise \ac{dp} baseline.
\end{itemize}
\begin{notation*}
For a function $F(x,y)$, the notation $\nabla F(x;y)\triangleq\frac{\partial}{\partial x}F(x,y)$
denotes the partial derivative of the function $F$ with respect to
variable $x$, with $y$ held as a fixed parameter.
\end{notation*}

\section{System Model}

To easily illustrate the connection between location privacy and radio
map construction, we consider a virtual environment embedded radio
map model, which has a set of variables that explicitly capture the
environment properties. We consider a dense urban environment with
$N$ users distributed on the ground and are served by \acpl{bs}
deployed either on towers or carried by low-altitude aerial platforms.
Note that in such an environment, it is highly possible that the link
between the \ac{bs} and mobile user is blocked by buildings and trees,
depending on the relative locations of the nodes. This motivates the
construction of radio maps to capture the link quality between an
arbitrary mobile location $\mathbf{p}_{\textrm{u}}\in\mathbb{R}^{3}$
and an arbitrary aerial \ac{bs} location $\mathbf{p}_{\textrm{d}}\in\mathbb{R}^{3}$.

\subsection{A Virtual Environment Embedded Radio Map Model}

Denote the communication link $\mathbf{p}=(\mathbf{p}_{\textrm{u}},\mathbf{p}_{\textrm{d}})$
as the location pair of the ground user at $\mathbf{p}_{\textrm{u}}$
and the \ac{bs} at $\mathbf{p}_{\textrm{d}}$. A radio map $\gamma(\mathbf{p};\mathbf{h})$
is defined as a mapping from the location pair $\mathbf{p}$ to the
corresponding channel gain $\gamma$ between the two locations $\mathbf{p}_{\text{u}}$
and $\mathbf{p}_{\textrm{d}}$, where $\mathbf{h}$ is the environment
parameter.

\subsubsection{Segmented Propagation Model}

Conventional statistical channel model predicts the channel gain based
on the propagation conditions, which can be modeled based on the environment
parameter $\mathbf{h}$ as to be illustrated later. Let the distance
between the ground user and the aerial \ac{bs} communication be $d(\mathbf{p})=\|\mathbf{p}_{\text{u}}-\mathbf{p}_{\text{d}}\|_{2}$.
Denote $\mathcal{D}_{0}$ as the set of location pairs $\mathbf{p}$
where the link between the user and the aerial \ac{bs} is in \ac{los},
and $\mathcal{D}_{1}$ as the set of location pairs where the link
is in \ac{nlos}. Thus, the channel gain for \ac{los} conditions
is modeled by $\beta_{0}+\alpha_{0}\log_{10}d(\mathbf{p})\text{, for }\mathbf{p}\in\mathcal{D}_{0}$,
and for \ac{nlos} conditions, it is given by $\beta_{1}+\alpha_{1}\log_{10}d(\mathbf{p})\text{, for }\mathbf{p}\in\mathcal{D}_{1}$.
The parameters are to be learned in the radio map construction process.

\subsubsection{Virtual Environment Model}

The challenge is to characterize the shape of the propagation regions
$\mathcal{D}_{0}$ and $\mathcal{D}_{1}$. We adopt an equivalent
virtual environment model with a simplified ray-tracing mechanism.
The general idea is to introduce virtual obstacles at specific locations,
each with an appropriate height to possibly intersect with the direct
communication path to model the \ac{los} or \ac{nlos} scenario.

\begin{figure}
\begin{centering}
\includegraphics[width=0.8\columnwidth]{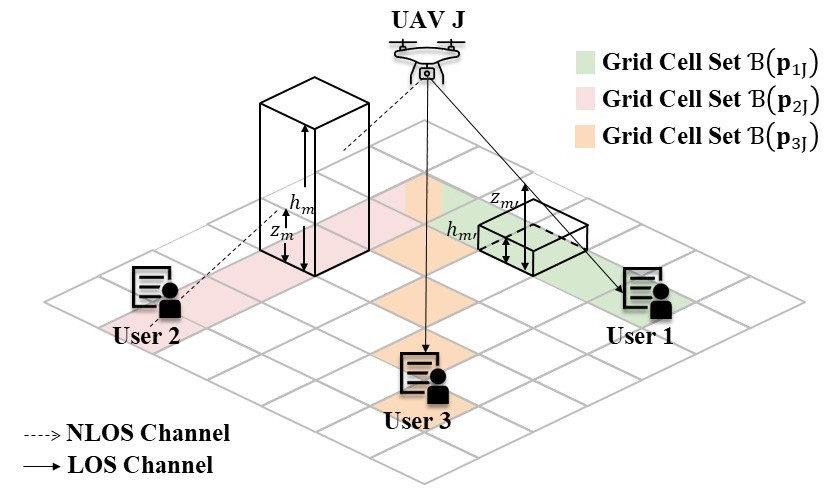}
\par\end{centering}
\caption{The schematic diagram of virtual obstacle model with the simplified
direct communication.}

\label{fig:virobs}
\end{figure}

Specifically, consider that the area is uniformly divided into $M$
grid cells, where each cell $m$ is possibly occupied by a virtual
obstacle with height $h_{m}$, $m=1,2,\dots,M$, where $h_{m}=0$
for no obstacle. Collectively, the heights of all virtual obstacles
constitute a virtual obstacle map, which can be represented by $\mathbf{h}\in\mathbb{R}^{M}$.
For a direct link $\mathbf{p}=(\mathbf{p}_{\text{u}},\mathbf{p}_{\text{d}})$
between the user and the aerial \ac{bs}, define $\mathcal{B}(\mathbf{p})$
as the set of grid cells underneath the link, and $z_{m}$ as the
corresponding link height at each cell $m\in\mathcal{B}(\mathbf{p})$.
The link is considered to be \ac{los} if $h_{m}<z_{m}$ for all $m\in\mathcal{B}(\mathbf{p})$,
indicating that none of the virtual obstacles blocks the link. Conversely,
\ac{nlos} condition occurs if $h_{m}\geq z_{m}$ for at least one
$m\in\mathcal{B}(\mathbf{p})$, indicating that the link $\mathbf{p}$
is blocked by at least one virtual obstacle. An illustration of the
relationship between $h_{m}$ and $z_{m}$ is shown in Fig. \ref{fig:virobs}.
Thus, the virtual obstacle model embedded in the channel gain model
can be expressed as

\begin{equation}
\gamma(\mathbf{p};\mathbf{h})=\begin{cases}
\beta_{0}+\alpha_{0}\log_{10}d(\mathbf{p}), & h_{m}<z_{m},\forall m\in\mathcal{B}(\mathbf{p})\\
\beta_{1}+\alpha_{1}\log_{10}d(\mathbf{p}), & h_{m}\geq z_{m},\exists m\in\mathcal{B}(\mathbf{p}).
\end{cases}\label{eq:seg-chan-gain}
\end{equation}

\subsubsection{Radio Map Model with Smoothed Transition}

However, the segmented function (\ref{eq:seg-chan-gain}) leads to
abrupt transitions between \ac{los} and \ac{nlos} regions which
may not be consistent with the reality. To address this, we propose
the use of continuous functions to create a smoother transition between
\ac{los} and \ac{nlos}. Consequently, the radio map is formulated
as
\begin{equation}
\begin{aligned}\gamma(\mathbf{p};\mathbf{h}) & =(\beta_{0}+\alpha_{0}\log_{10}d(\mathbf{p}))S(\mathbf{p};\mathbf{h})\\
 & \ \ +(\beta_{1}+\alpha_{1}\log_{10}d(\mathbf{p}))(1-S(\mathbf{p};\mathbf{h}))
\end{aligned}
\label{eq:radio map}
\end{equation}
where $S(\mathbf{p};\mathbf{h})$ is a sigmoid function to approximately
model the impact of the \ac{los} and \ac{nlos} propagation based
on the heights of virtual obstacles $\mathbf{h}$. Specifically, the
function $S(\mathbf{p};\mathbf{h})$ is defined as 
\begin{equation}
S(\mathbf{p},\mathbf{h})=\prod_{m\in\mathcal{B}(\mathbf{p})}\psi(z_{m}(\mathbf{p})-h_{m})\label{eq:virtual-obsatcle}
\end{equation}
where $\psi(x)$ is given by $(1+e^{-x})^{-1}$. This function represents
a logistic function with a smooth transition between $0$ and $1$.
If virtual obstacles in all passing grid cells do not obscure the
direct path $\mathbf{p}$, i.e., $z_{m}(\mathbf{p})>h_{m},\forall m\in\mathcal{B}(\mathbf{p})$,
then $S(\mathbf{p},\mathbf{h})\approx1$ and $\gamma(\mathbf{p};\mathbf{h})$
approximately equals to the first expression in (\ref{eq:seg-chan-gain}),
indicating an \ac{los} scenario; on the other hand, if there exists
a virtual obstacle in the passing grid cell that obscures the communication
link, i.e., $z_{m}(\mathbf{p})<h_{m},\exists m\in\mathcal{B}(\mathbf{p})$,
then $S(\mathbf{p},\mathbf{h})\approx0$ and $\gamma(\mathbf{p};\mathbf{h})$
approximately equals to the second expression in (\ref{eq:seg-chan-gain}),
indicating an \ac{nlos} scenario.

Based on (\ref{eq:radio map}), the \ac{rss} from the aerial node
$\mathbf{p}_{\text{d}}$ measured at the user location $\mathbf{p}_{\text{u}}$
in a logarithmic scale is formulated as
\begin{equation}
y=\gamma(\mathbf{p};\mathbf{h}^{*})+\xi\label{eq:observed RSS}
\end{equation}
where $\xi$ captures the randomness due to the residual shadowing
and the measurement noise.

\subsection{Radio Map Construction via Federated Learning}

Consider a collaborative data collection approach, where each user
$i$ collects channel measurements $\gamma$ following (\ref{eq:observed RSS})
with \acpl{bs} at various locations. The user-\ac{bs} location pair
dataset collected by user $i$ is denoted by $\mathcal{N}_{i}$.

If all the data were available at a central server, the radio map
(\ref{eq:radio map}) can be constructed by estimating the heights
$\mathbf{h}$ and the propagation parameters $\boldsymbol{\theta}=\{\alpha_{0},\beta_{0},\alpha_{1},\beta_{1}\}$
via minimizing the least-squares cost 
\begin{equation}
F(\boldsymbol{\theta},\mathbf{h})\triangleq\frac{1}{J}\sum_{i=1}^{N}\sum_{j\in\mathcal{N}_{i}}[y_{j}-\gamma(\mathbf{p}_{j};\boldsymbol{\theta},\mathbf{h})]^{2}\label{eq:objective}
\end{equation}
where $\mathbf{p}_{j}$ is a sample in the measurement dataset, $J=\sum_{i=1}^{N}J_{i}$
denotes the total size of global dataset from all $N$ users, and
$J_{i}=|\mathcal{N}_{i}|$ denotes the size of $i$th user dataset
$\mathcal{N}_{i}$.

We consider a scenario where the measurement data from each user are
kept locally and not shared, while \ac{fel} is used for radio map
construction. Specifically, define 
\begin{equation}
F_{i}(\boldsymbol{\theta},\mathbf{h})=\frac{1}{J_{i}}\sum_{j\in\mathcal{N}_{i}}[y_{j}-\gamma(\mathbf{p}_{j};\boldsymbol{\theta},\mathbf{h})]^{2}\label{eq:local loss function}
\end{equation}
as the least-squares cost function of the $i$th user. The goal cost
function (\ref{eq:objective}) relates to the individual cost function
as
\begin{equation}
F(\boldsymbol{\theta},\mathbf{h})=\sum_{i=1}^{N}\frac{J_{i}}{J}F_{i}(\boldsymbol{\theta},\mathbf{h}).\label{eq:aggregation}
\end{equation}

\begin{figure}
\begin{centering}
\includegraphics[width=0.95\columnwidth]{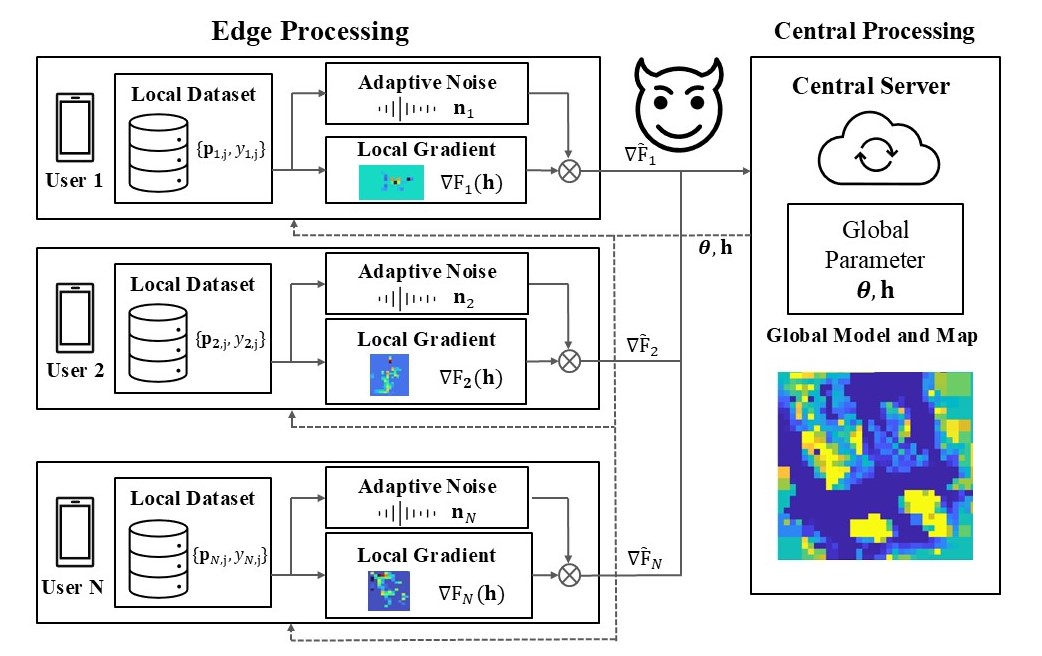}
\par\end{centering}
\caption{The \ac{fel} framework for model training: Users calculate local
gradients on a locally decomposed model, add artificial noise to the
gradient and upload them to the central server for aggregation.}

\label{fig:Fed-framwork}
\end{figure}

As illustrated in Fig.~\ref{fig:Fed-framwork}, the \ac{fel} radio
map construction process consists of the following iterative steps:
\begin{enumerate}
\item \textbf{Broadcasting}: In the $t$th communication round, the central
server broadcasts the current global virtual obstacle map $\mathbf{h}(t)$
and the global propagation parameters $\boldsymbol{\theta}(t)$ to
all participating users.
\item \textbf{Local Training}: Upon receiving the global models, each user
$i$ computes the local gradients based on its local dataset $\mathcal{N}_{i}$:
\begin{itemize}
\item The partial gradient for the environment map $\mathbf{h}$
\begin{equation}
\nabla F_{i}(\mathbf{h}(t);\boldsymbol{\theta})=\frac{1}{J_{i}}\sum_{j\in\mathcal{N}_{i}}\nabla[y_{j}-\gamma(\mathbf{p}_{j};\mathbf{h}(t))]^{2}\label{eq:local update}
\end{equation}
\item The partial gradient for propagation parameters $\boldsymbol{\theta}$
\begin{equation}
\nabla F_{i}(\boldsymbol{\theta}(t);\mathbf{h})=\frac{1}{J_{i}}\sum_{j\in\mathcal{N}_{i}}\nabla[y_{j}-\gamma(\mathbf{p}_{j};\boldsymbol{\theta}(t))]^{2}
\end{equation}
\end{itemize}
\item \textbf{Model Aggregation}: The central server aggregates the gradients
received from $N$ users to update the global environment parameter
as
\begin{equation}
\mathbf{h}(t+1)=\mathbf{h}(t)-\eta_{\text{h}}\sum_{i=1}^{N}\frac{J_{i}}{J}\nabla F_{i}(\mathbf{h}(t);\boldsymbol{\theta})\label{eq:h aggregation}
\end{equation}
where $\eta_{\text{h}}$ is the learning rate for $\mathbf{h}$, and
update the global propagation parameter as
\begin{equation}
\boldsymbol{\theta}(t+1)=\boldsymbol{\theta}(t)-\eta_{\theta}\sum_{i=1}^{N}\frac{J_{i}}{J}\nabla F_{i}(\boldsymbol{\theta}(t);\mathbf{h})\label{eq:theta aggregation}
\end{equation}
where $\eta_{\theta}$ is the learning rate for $\boldsymbol{\theta}$.
\end{enumerate}
Under mild assumptions, the \ac{fel} procedure is guaranteed to converge\cite{li2019convergence},
and the converged solution corresponds to a stationary point of the
objective function in (\ref{eq:objective}). However, although \ac{fel}
does not require data sharing, there is still location privacy leakage
as analyzed in the next section.

\section{Position Privacy Leakage in Conventional Federated Learning}

In this section, we analyze how the location information may leak
from the local gradient in radio map construction, where we identify
an adversary strategy that can infer user location with high precision
from the uploaded local gradients.

\subsection{Statistics of Local Gradients\label{subsec:Statistics-of-Local}}

It can be shown that the second-order statistics of the local gradient
of the environment parameter $\mathbf{h}$ has a strong correlation
with the location of the user. Specifically, for the environment parameter
$h_{m}$ that is geographically closer to the user, a larger magnitude
of the local gradients is expected.

Let $G_{i,m}$ denote the $m$th element of the local gradient vector
$\nabla F_{i}(\mathbf{h})$, which represents the gradient contribution
with respect to the height $h_{m}$ of the $m$th virtual obstacle
for user $i$. From the loss function defined in (\ref{eq:local loss function}),
we can derive the expression for $G_{i,m}$ as
\begin{equation}
G_{i,m}=\frac{2}{J_{i}}\sum_{j\in\mathcal{N}_{i}}\Gamma(\mathbf{p}_{j};\mathbf{h})D(\mathbf{p}_{j};\mathbf{h},m)\label{eq:Gim_decomposed}
\end{equation}
where $\Gamma(\mathbf{p}_{j};\mathbf{h})$ is given by
\begin{equation}
\begin{aligned}\Gamma(\mathbf{p}_{j};\mathbf{h}) & =(y_{j}-\gamma(\mathbf{p}_{j};\mathbf{h}))\\
 & \enspace\times[(\beta_{1}+\alpha_{1}\log_{10}d(\mathbf{p}_{j}))-(\beta_{0}+\alpha_{0}\log_{10}d(\mathbf{p}_{j}))]
\end{aligned}
\label{eq:g_i,m first term}
\end{equation}
which captures the model prediction error scaled by the difference
between the \ac{nlos} and \ac{los} path loss models, and $D(\mathbf{p}_{j};\mathbf{h},m)$
is given by
\begin{equation}
D(\mathbf{p}_{j};\mathbf{h},m)=\frac{\partial}{\partial h_{m}}S(\mathbf{p}_{j};\mathbf{h})\label{eq:g_i,m second term}
\end{equation}
which measures how sensitive the smoothed \ac{los} function $S(\mathbf{p}_{j};\mathbf{h})$
is to the change in the height $h_{m}$ of the specific grid cell
$m$.

In addition, recall that $\mathcal{B}(\mathbf{p}_{j})$ represents
the collection of grid cells underneath by the communication link
$\mathbf{p}_{j}$. We define $\mathcal{N}_{i,m}\subseteq\mathcal{N}_{i}$
to be the subset of user $i$ measurements whose paths traverse cell
$m$ , i.e.,
\begin{equation}
\mathcal{N}_{i,m}=\{j|j\in\mathcal{N}_{i},m\in\mathcal{B}(\mathbf{p}_{j})\}.
\end{equation}
For the link not above the grid cell $m$, i.e. $j\notin\mathcal{N}_{i,m}$,
the change in the height $h_{m}$ does not affect the probability
function $S(\mathbf{p}_{j};\mathbf{h})$, rendering $D(\mathbf{p}_{j};\mathbf{h},m)=0$.
Thus, the gradient (\ref{eq:Gim_decomposed}) can be expressed as
\begin{equation}
\begin{aligned}G_{i,m} & =\frac{2}{J_{i}}\sum_{j\in\mathcal{N}_{i,m}}\Gamma(\mathbf{p}_{j};\mathbf{h})D(\mathbf{p}_{j};\mathbf{h},m)\\
 & \quad+\frac{2}{J_{i}}\sum_{j\notin\mathcal{N}_{i,m}}\Gamma(\mathbf{p}_{j};\mathbf{h})D(\mathbf{p}_{j};\mathbf{h},m)\\
 & =\frac{2}{J_{i}}\sum_{j\in\mathcal{N}_{i,m}}\Gamma(\mathbf{p}_{j};\mathbf{h})D(\mathbf{p}_{j};\mathbf{h},m).
\end{aligned}
\label{eq:simplification of local gradient element}
\end{equation}

Now we investigate the relationship between the magnitude of $G_{i,m}$
in (\ref{eq:simplification of local gradient element}) and the distance
away from the $i$th user. For the $i$th user at location $\mathbf{p}_{\text{u},i}$,
define a subset of grid cells $\mathcal{M}_{i}=\{m\geq0|\mathbf{c}_{m}=\mathbf{p}_{\text{u},i}+m\boldsymbol{\delta}\}$,
representing cells along an arbitrary direction $\boldsymbol{\delta}$
originating from the user. The following result establishes the expected
magnitude of the local gradient elements.
\begin{thm}[Gradient Attenuation with Distance\label{thm:gradient attenuation}]
If $h_{m}<h_{m}^{*},\forall m\in\mathcal{M}_{i}$ or $h_{m}>h_{m}^{*},\forall m\in\mathcal{M}_{i}$,
then, for any $m\in\mathcal{M}_{i}$ such that $\|\mathbf{c}_{m-1}-\mathbf{p}_{\text{u},i}\|<\|\mathbf{c}_{m}-\mathbf{p}_{\text{u},i}\|$,
the statistics of the corresponding local gradients satisfy 
\begin{equation}
\mathbb{E}\{G_{i,m-1}^{2}\}\geq\mathbb{E}\{G_{i,m}^{2}\}.
\end{equation}
\end{thm}
\begin{proof}
See Appendix \ref{sec:Proof-of-Theorem1}.
\end{proof}
Theorem \ref{thm:gradient attenuation} implies that, along any given
direction from the user, the expected squared magnitude of the decomposed
local gradient $G_{i,m}$ tends to decrease as the distance from the
user to the grid cell $m$ increases, provided the estimated obstacle
heights $h_{m}$ along that direction are consistently higher or lower
than their true values $h_{m}^{*}$. As a result, an adversary may
exploit this property to infer the location of the user.

\subsection{An Adversary Localization Strategy\label{subsec:An-Adversary-Localization}}

Inspired by Theorem \ref{thm:gradient attenuation}, we demonstrate
an adversary strategy to infer the user location based on the local
gradients under the \ac{fel} framework. Specifically, we may adopt
a \ac{wcl} approach based on the local gradients $G_{i,m}$ as follows.
\begin{equation}
\hat{\mathbf{p}}_{\text{u},i}=\dfrac{\sum_{m=1}^{M}|G_{i,m}|^{\nu}\mathbf{c}_{m}}{\sum_{m=1}^{M}|G_{i,m}|^{\nu}}\label{eq:wcl-localization}
\end{equation}
where $\mathbf{c}_{m}$ is the coordinate of the $m$th grid cell,
$|G_{i,m}|$ is the magnitude of the local gradient component from
user $i$ for cell $m$, and $\nu$ is a positive parameter.
\begin{table}
\caption{The \ac{rmse} of the proposed adversary localization strategy \label{tab:The-localization-mae-pure}}

\centering{}\centerline{%
\begin{tabular*}{1\columnwidth}{@{\extracolsep{\fill}}cccccc}
\toprule 
Parameter $\nu$ & 1 & 2 & 5 & 10 & Inf\tabularnewline
\midrule
\midrule 
$1$st epoch (m) & 7.25 & 6.90 & 8.26 & 8.61 & 9.00\tabularnewline
\midrule 
$100$th epoch (m) & 20.69 & 23.42 & 17.67 & 17.23 & 18.41\tabularnewline
\midrule 
$200$th epoch (m) & 27.44 & 28.93 & 30.27 & 30.00 & 31.44\tabularnewline
\midrule 
Average (m) & 22.10 & 24.32 & 23.52 & 23.86 & 24.24\tabularnewline
\bottomrule
\end{tabular*}}
\end{table}

Table~\ref{tab:The-localization-mae-pure} summarizes the location
\ac{rmse} of the \ac{wcl} adversary using local gradients obtained
at different training epochs. The results show that a localization
error below $10$ meters can be obtained in the early stages of training,
implying a significant privacy risk. This risk is persistent over
different parameter settings and training epochs. The most severe
leakage occurs in the first epoch, where the adversary achieves a
localization error as low as $7\lyxmathsym{\textendash}9$ meters.
This observation directly aligns with Theorem~\ref{thm:gradient attenuation},
as the training was initialized by $h_{m}=H_{\max}$, an optimization
strategy that was shown to be theoretically desired in previous studies
\cite{10041012}. The accuracy of the adversary strategy degrades
as the training process evolves, but an accuracy of less than 30 m
can be achieved under $\nu=1,2$, implying a significant location
privacy leakage for the entire \ac{fel} training process. This suggests
that the gradient remains strongly geographically correlated even
beyond the conditions of the Theorem~\ref{thm:gradient attenuation}.

\section{A Remedy via Geometry-Aligned Differential Privacy\label{sec:A-Remedy-via}}

In the previous section, it is revealed that the local gradients under
the \ac{fel} framework for radio map construction still contain location
information of the user that can be recovered by the central server.
The fundamental reason, as illustrated in Theorem~\ref{thm:gradient attenuation}
based on the radio map model (\ref{eq:radio map}), is that some of
the model parameters, such as $\mathbf{h}$ in (\ref{eq:radio map}),
contains environment information and consequently, there may exist
some adversary strategies to infer the user location based on the
gradients of those environment-related model parameters.

To mitigate the location privacy leakage, we consider a \ac{dp} approach
that adds artificial noise to the local gradient of the obstacle maps
$\nabla F_{i}(\mathbf{h}(t);\boldsymbol{\theta}(t))$ in (\ref{eq:local update})
to obscure the location information implicitly conveyed by the environment-related
model parameters. For clarity of discussion, we define the original
local gradient of the obstacle map as 
\begin{equation}
\mathbf{g}_{i}(t)\triangleq\nabla F_{i}(\mathbf{h}(t);\boldsymbol{\theta}(t)).
\end{equation}
In the \ac{dp} approach\cite{wei2020federated}, it is first clipped
as
\begin{equation}
\bar{\mathbf{g}}_{i}(t)=\mathbf{g}_{i}(t)\cdot\min\left(1,\frac{C}{\|\mathbf{g}_{i}(t)\|_{2}}\right)\label{eq:clipped gradient}
\end{equation}
to ensure that $\|\bar{\mathbf{g}}_{i}(t)\|_{2}\leq C$, where $C$
is the clipping threshold for bounding $\mathbf{g}_{i}(t)$. Then
noise is added to the clipped local gradient $\bar{\mathbf{g}}_{i}(t)$
to obtain the uploaded noisy version 
\begin{equation}
\tilde{\mathbf{g}}_{i}(t)=\bar{\mathbf{g}}_{i}(t)+\mathbf{n}_{i}(t)\label{eq:noisy gradient}
\end{equation}
where $\mathbf{n}_{i}(t)$ is zero-mean Gaussian noise $n_{i,m}\sim(0,\sigma_{i,m}^{2})$,
where the component-wise variance $\sigma_{i,m}^{2}$ is to be determined
later in this section.

To control the overall privacy-utility trade-off, we introduce a design
parameter $\mu\geq0$ that serves as a noise budget. This parameter
bounds the total noise variance relative to the signal strength, according
to the following condition
\begin{equation}
\mathbb{E}\{\|\mathbf{n}_{i}(t)\|^{2}\}\leq\mu\|\bar{\mathbf{g}}_{i}(t)\|^{2}.\label{eq:noise budget}
\end{equation}

While adding noise deteriorates the convergence of \ac{fel}, the
key challenge here is to strike a balance between learning performance
and location privacy protection through controlling the noise variance
$\sigma_{i,m}^{2}$ under the budget set by $\mu$. In this section,
we first establish the convergence of the federated radio map construction
under a given noise budget. Then, we develop a mechanism to control
the noise variance $\sigma_{i,m}^{2}$ of different model parameters
to maximize location protection.

\subsection{Convergence under Differential Privacy}

We first analyze the impact of noise on the \ac{fel} global objective
function $F(\boldsymbol{\theta},\mathbf{h})$. Existing \ac{fel}
convergence analyses assume identical noise variance. It is not known
whether a similar convergence result holds if the distribution of
the gradient noise $n_{i,m}$ is not identical.

First, by examining the formulation (\ref{eq:radio map}), (\ref{eq:virtual-obsatcle}),
(\ref{eq:objective}), (\ref{eq:aggregation}), one can conclude that
$F_{i}(\boldsymbol{\theta},\mathbf{h})$ and $F(\boldsymbol{\theta},\mathbf{h})$
are Lipschitz continuous since they are compositions of smooth functions.
Denote the Lipschitz constant as $L$, i.e.,
\begin{equation}
\|\nabla F(\mathbf{x}_{1})-\nabla F(\mathbf{x}_{2})\|\leq L\|\mathbf{x}_{1}-\mathbf{x}_{2}\|\label{eq:L-smooth}
\end{equation}
where $\mathbf{x}=(\boldsymbol{\theta},\mathbf{h})$. We further make
the following assumption on the local gradient $\mathbf{g}_{i}(t)=\nabla F_{i}(\mathbf{h}(t);\boldsymbol{\theta}(t))$.
\begin{assumption}[$B$-Local Dissimilarity]
The local gradient $\mathbf{g}_{i}(t)=\nabla F_{i}(\mathbf{h}(t))$
is $B$-locally dissimilar at a given parameter $\mathbf{h}$ as 
\begin{equation}
\mathbb{E}\{\|\mathbf{g}_{i}(t)\|^{2}\}\leq B^{2}\mathbb{E}\{\|\mathbf{g}(t)\|^{2}\}
\end{equation}
where $\mathbf{g}(t)=\sum_{i=1}^{N}\frac{J_{i}}{J}\mathbf{g}_{i}(t)$
according to (\ref{eq:objective})\textendash (\ref{eq:local loss function}),
and $B$ is a finite constant that quantifies the degree of deviation
between local and global gradients.
\end{assumption}
This condition holds in our model because the continuous differentiability
of our objective function (\ref{eq:objective}) ensures that both
local gradients $\mathbf{g}_{i}(t)$ and the global gradient $\mathbf{g}(t)$
are bounded within any compact set of parameters.

Let $\tilde{J}=\sum_{i=1}^{N}J_{i}^{2}/J^{2}$ denote the heterogeneity
of the size of the local datasets. From Cauchy-Schwartz inequality,
we have $\frac{1}{N}\leq\tilde{J}<1$, where the lower bound is achieved
when $J_{1}=J_{2}=\cdots J_{N}$, and $\tilde{J}$ approaches to $1$
when $J_{i}$ are highly heterogeneous.
\begin{lem}[Global Loss Reduction Bound\label{lemma:loss_reduction_bound}]
 Given the propagation parameter $\boldsymbol{\theta}(t)$ in each
training round, the expected decrease of the loss is bounded as
\begin{equation}
\mathbb{E}\{F(\mathbf{h}(t+1),\boldsymbol{\theta}(t))-F(\mathbf{h}(t),\boldsymbol{\theta}(t))\}\le-\kappa_{h}\mathbb{E}\{\|\mathbf{g}(t)\|_{2}^{2}\}+\kappa_{0}\label{eq:lemma_h}
\end{equation}
where $\kappa_{h}=\eta_{h}\left(\frac{1}{2}-\frac{L\eta_{h}}{2}B^{2}(1+\mu\tilde{J})\right)$,
and $\kappa_{0}=\frac{\eta_{h}B^{4}L^{4}}{8C^{2}}$.

Given the obstacle map $\mathbf{h}(t+1)$ in each training round,
the expected decrease of the loss is bounded as
\begin{equation}
\begin{aligned}\mathbb{E}\{F(\mathbf{h}(t+1),\boldsymbol{\theta}(t+1))-F(\mathbf{h}(t+1),\boldsymbol{\theta}(t))\}\\
\quad\le-\kappa_{\theta}\mathbb{E}\{\|\nabla F(\boldsymbol{\theta}(t);\mathbf{h}(t+1))\|_{2}^{2}\}
\end{aligned}
\label{eq:lemma_theta}
\end{equation}
where $\kappa_{\theta}=\eta_{\theta}-\frac{L\eta_{\theta}^{2}}{2}$.
\end{lem}
\begin{proof}
See Appendix \ref{sec:Proof-of-Lemma1}.
\end{proof}
Combining (\ref{eq:lemma_h}) and (\ref{eq:lemma_theta}), the expected
decrease of the global function is formulated as
\begin{equation}
\begin{aligned}\mathbb{E}\{F(\mathbf{h}(t+1),\boldsymbol{\theta}(t+1))-F(\mathbf{h}(t),\boldsymbol{\theta}(t))\}\quad\quad\quad\quad\quad\quad\\
\le-\kappa_{h}\mathbb{E}\{\|\mathbf{g}(t)\|_{2}^{2}\}+\kappa_{0}-\kappa_{\theta}\mathbb{E}\{\|\nabla F(\boldsymbol{\theta}(t))\|_{2}^{2}\}.
\end{aligned}
\label{eq:lemma1_expected_decreace}
\end{equation}
By iteratively applying (\ref{eq:lemma1_expected_decreace}) from
$t=0$ to $t=T-1$, we obtain a bound on the average of the squared
gradient norms 
\begin{equation}
\begin{aligned}\frac{1}{T}\sum_{t=0}^{T-1}\kappa(\mathbb{E}\{\|\mathbf{g}(t)\|_{2}^{2}\}+\mathbb{E}\{\|\nabla F(\boldsymbol{\theta}(t))\|_{2}^{2}\})\quad\\
\le\frac{1}{T}\mathbb{E}\{F(\mathbf{h}(0),\boldsymbol{\theta}(0))-F(\mathbf{h}(T),\boldsymbol{\theta}(T))\}+\kappa_{0}
\end{aligned}
\label{eq:cummulative-loss}
\end{equation}
where $\kappa=\min(\kappa_{h},\kappa_{\theta})$.

Since the cost function $F(\mathbf{h},\boldsymbol{\theta})$ is non-negative,
as the number of iterations $T$ tends to $\infty$, the first term
on the right hand side tends to $0$. Recall that the total gradient
is structured as 
\begin{equation}
\nabla F(\mathbf{h}(t),\boldsymbol{\theta}(t))=[\mathbf{g}(t)^{\text{T}},\nabla F(\boldsymbol{\theta}(t))^{\text{T}}]^{\text{T}}.
\end{equation}
As a result, if $\kappa>0$, we obtain
\begin{equation}
\begin{aligned}\frac{1}{T}\sum_{t=0}^{T-1}\mathbb{E}\{\|\nabla F(\mathbf{h}(t),\boldsymbol{\theta}(t))\|_{2}^{2}\}\quad\quad\quad\quad\quad\quad\quad\quad\quad\quad\quad\\
\leq\frac{1}{T\kappa}\mathbb{E}\{F(\mathbf{h}(0),\boldsymbol{\theta}(0))-F(\mathbf{h}(T),\boldsymbol{\theta}(T))\}+\frac{\eta_{h}B^{4}L^{4}}{8C^{2}\kappa}
\end{aligned}
\label{eq:initial bound}
\end{equation}
where the first term tends to $0$ as $T\to\infty$, and the second
term is a small number that also tends to zero as the clipping number
$C\to\infty$, i.e., no clipping. Building on this, Corollary~\ref{cor:Convergence}
provides an upper bound on the expected norm of the total gradient
under heterogeneous noise conditions.
\begin{cor}[Convergence under Heterogeneous Noise\label{cor:Convergence}]
If the learning rate satisfies $0<\eta_{h}<\frac{1}{LB^{2}(1+\mu\tilde{J})}$
and $0<\eta_{\theta}<\frac{2}{L}$, then the total gradient is upper
bounded by
\begin{equation}
\lim_{t\to\infty}\ \mathbb{E}\{\|\nabla F(\mathbf{h}(t),\boldsymbol{\theta}(t))\|_{2}\}\leq\sqrt{\frac{\eta_{h}B^{4}L^{4}}{8C^{2}\kappa}}
\end{equation}
which tends to 0 as the clipping number $C\to\infty$.
\end{cor}
\begin{proof}
See Appendix \ref{sec:Proof-of-Corollary}.
\end{proof}
Corollary~\ref{cor:Convergence} establishes that the federated radio
map construction algorithm attains a bounded expected norm for the
total gradient, even in the presence of heterogeneous artificial noise
added to the local gradients as in (\ref{eq:noisy gradient}).

Lemma~\ref{lemma:loss_reduction_bound}, on the other hand, implies
a tradeoff between reducing the expected gradient norm and preserving
privacy. Maximizing the descent rate $\kappa_{h}$ leads to faster
loss reduction with respect to $\mathbf{h}$. Under this goal, the
maximum learning rate for a given noise budget $\mu$ is 
\begin{equation}
\eta_{h}^{*}=\frac{1}{LB^{2}(1+\mu\tilde{J})}\label{eq:maximum-leanring-rate}
\end{equation}
where a larger $\mu$ reduces $\eta_{h}^{*}$, limiting the convergence
speed. Conversely, for any fixed learning rate $\eta_{h}>0$, ensuring
$\kappa_{h}>0$ in (\ref{eq:lemma1_expected_decreace}) requires the
noise budget to be bounded by
\begin{equation}
\mu<\frac{1}{\tilde{J}}\big(\frac{1}{B^{2}L\eta_{h}}-1\big).\label{eq:noise-budget}
\end{equation}

These expressions quantify the trade-off between privacy and learning
performance. A larger noise budget $\mu$ improves privacy but reduces
the allowable learning rate. As $\mu$ increases, both $\eta_{h}$
and the descent rate $\kappa_{h}$ decrease. This slows the update
of $\mathbf{h}$ and degrades the efficiency of the alternating optimization
between $\mathbf{h}$ and $\boldsymbol{\theta}$.

Furthermore, the convergence bound in Corollary~\ref{cor:Convergence}
includes a residual term controlled by the gradient clipping threshold
$C$, which is introduced to mitigate heterogeneous noise effects.
A smaller $C$ improves robustness to noise and strengthens privacy.
However, it increases the residual $\sqrt{\frac{\eta_{h}B^{4}L^{4}}{8C^{2}\kappa}}$
leading to a looser convergence bound. In contrast, letting $C\to\infty$
eliminates the residual but weakens noise control and may reduce privacy
protection.

\subsection{Principles of the Geometry-Aligned Differential Privacy}

While the total noise budget $\mu$ affects the convergence bound
as stated in Lemma~\ref{lemma:loss_reduction_bound} and Corollary~\ref{cor:Convergence},
we have the flexibility to distribute the total noise budget $\mu$
over individual grid cells $\sigma_{i,m}^{2}$ to enhance privacy
protection without affecting the convergence bound.

Denote $\tilde{G}_{i,m}=\bar{G}_{i,m}+n_{i,m}$ as the $m$th element
of the noisy gradient element on the $i$th user. The vector of noise
energies across all grid cells is defined as $\bm{\sigma}_{i}=(\sigma_{i,1}^{2},...,\sigma_{i,M}^{2})$.
We propose a geometry-aligned noise allocation scheme for $\bm{\sigma}_{i}$
to achieve two privacy objectives. First, we aim at maximizing the
numerical uncertainty of the estimated user location by the adversary,
which is quantified by the expected \ac{mse} between the estimated
location $\tilde{\mathbf{p}}_{\text{u},i}$ and the true location
$\mathbf{p}_{\text{u},i}$
\begin{equation}
P(\bm{\sigma}_{i})\triangleq\mathbb{E}\{\|\tilde{\mathbf{p}}_{\text{u},i}-\mathbf{p}_{\text{u},i}\|_{2}^{2}\}\label{eq:numerical confusion}
\end{equation}
where $\tilde{\mathbf{p}}_{\text{u},i}$ is the location estimate
that is probably computed by the adversary. Based on our location
leakage study in Table~\ref{tab:The-localization-mae-pure} for the
adversary strategy (\ref{eq:wcl-localization}), we simply pick the
form
\begin{equation}
\tilde{\mathbf{p}}_{\text{u},i}=\frac{\sum_{m=1}^{M}\tilde{G}_{i,m}^{2}\mathbf{c}_{m}}{\sum_{m=1}^{M}\tilde{G}_{i,m}^{2}}.\label{eq:WCL}
\end{equation}

\begin{figure}
\centerline{\includegraphics[width=1\columnwidth]{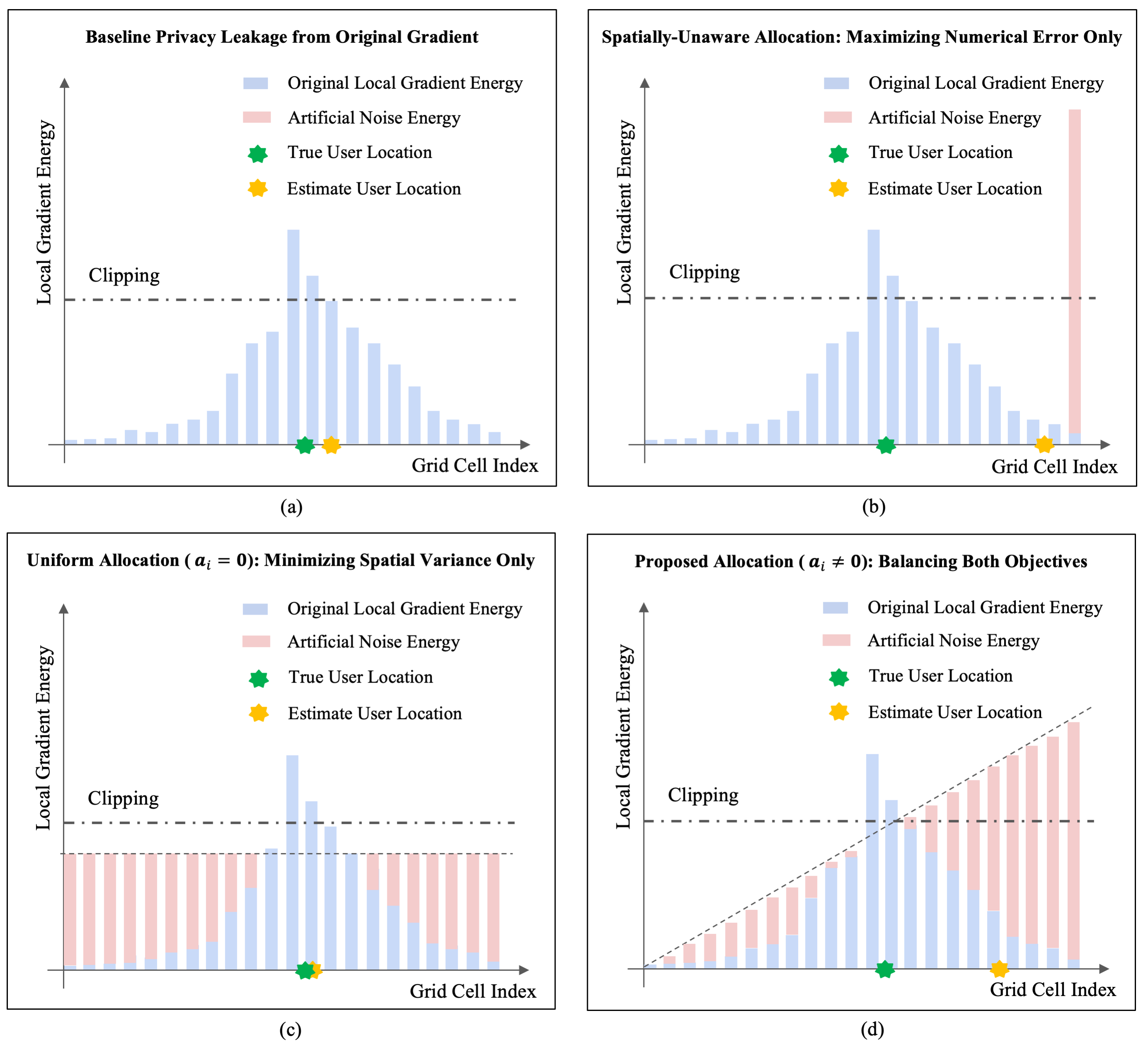}}

\caption{Illustration of the leaked location by the adversary strategy (\ref{eq:wcl-localization}):
(a) The original local gradient energy peaks near the true user location,
leading to a significant privacy leakage under a WCL attack. (b) A
spatially-unaware strategy that maximizes the localization error by
placing all noise on the edge fails to obfuscate the gradient's spatial
pattern. (c) A uniform noise allocation ($\mathbf{a}_{i}=\mathbf{0}$)
perfectly masks the spatial pattern but provides weak numerical protection
against the adversary. (d) The proposed spatially-aware linear allocation
($\mathbf{a}_{i}\protect\neq\mathbf{0}$) effectively balances both
privacy objectives by simultaneously creating a large localization
bias and masking the original gradient's shape.}

\label{fig:Linear solution}
\end{figure}

Second, we also need to obfuscate the spatial pattern of the noise
corrupted gradient $\tilde{G}_{i,m}$. An example is illustrated in
Fig.~\ref{fig:Linear solution}, where in Fig.~\ref{fig:Linear solution}(b),
if one only focused on the numerical confusion metric (\ref{eq:numerical confusion}),
the optimal choice will be allocating the entire noise budget to the
grid cells on the edge. On the contrary, we hope to distribute noise
smoothly over a large number of grid cells to cover the spatial pattern
of the gradient as shown in Fig.~\ref{fig:Linear solution}(d).

Toward this end, we define a new metric, \emph{spatial variance},
for the spatial smoothness of the noise corrupted gradient.
\begin{defn}[Spatial Variance]
The spatial variance across grid cells $m\in\{1,2,...,M\}$ is defined
as the expectation of the sample variance over this set of random
variables $\{\tilde{G}_{i,m}^{2}\}$
\begin{equation}
V(\bm{\sigma}_{i})\triangleq\mathbb{E}\Big\{\frac{1}{M}\sum_{m=1}^{M}\Big(\tilde{G}_{i,m}^{2}-\frac{1}{M}\sum_{n=1}^{M}\tilde{G}_{i,n}^{2}\Big)^{2}\Big\}.\label{eq:spatial variance-1}
\end{equation}
\end{defn}
It follows that if $\tilde{G}_{i,m}^{2}$ are spatially and statistically
identical, the spatial variance is minimized as 0, achieving the best
protection in terms of hiding the spatial pattern. Note that, this
extreme spatial smoothness may require a large noise budget or a small
gradient clipping parameter $C$, which significantly affects the
convergence rate.

To ease the design and achieve a good trade-off between location privacy
protection and convergence, we consider a spatially linear noise allocation
pattern with slope $\mathbf{a}_{i}$ and offset $b_{i}$. As a result,
the square of the noise corrupted gradient $\tilde{G}_{i,m}^{2}$
roughly distributed above $\mathbf{a}_{i}^{T}\mathbf{c}_{m}+b_{i}$,
where recall that $\mathbf{c}_{m}$ is the coordinate of the $m$th
grid cell.

Two numerical examples are given in Fig.~\ref{fig:Linear solution},
where the original clipped squared gradient $\bar{G}_{i,m}^{2}$ are
statistically centered near the user location (See Theorem~\ref{thm:gradient attenuation}).
In Fig.~\ref{fig:Linear solution}(c), adding noise following slope
$\mathbf{a}_{i}=\mathbf{0}$ can completely mask the spatial pattern
of the gradient, achieving a spatial variance of 0 according to (\ref{eq:spatial variance-1}),
but the location privacy protection is weak against adversary strategy
(\ref{eq:WCL}). In Fig.~\ref{fig:Linear solution}(d), the noise
is allocated such that the expected squared of the noise corrupted
gradient $\tilde{G}_{i,m}$ are roughly masked by $\mathbf{a}_{i}^{T}\mathbf{c}_{m}+b_{i}$.
Although the non-zero slope $\mathbf{a}_{i}$ leads to a larger spatial
variance under (\ref{eq:spatial variance-1}), it also contributes
to location estimation bias under adversary strategy (\ref{eq:WCL}),
achieving a better location privacy protection. Note that one can
also clip the gradient to help bring down the spatial variance under
the same total noise budget, at the cost of sacrificing the learning
performance as indicated by Corollary~\ref{cor:Convergence}.

Based on the above two design objectives, the noise budget for the
$m$th grid cell is given as
\begin{equation}
\sigma_{i,m}^{2}=\max\{0,\mathbf{a}_{i}^{T}\mathbf{c}_{m}+b_{i}-\bar{G}_{i,m}^{2}\}.
\end{equation}
The slope $\mathbf{a}_{i}$ and the offset $b_{i}$ are optimized
to balance the numeric localization error (\ref{eq:numerical confusion})
under the adversary strategy (\ref{eq:WCL}) and the spatial variance
(\ref{eq:spatial variance-1}) that hides the spatial pattern of the
gradient as

\begin{align}
\underset{\mathbf{a}_{i},b_{i}}{\text{maximize}} & \quad\quad P(\bm{\sigma}_{i})-\rho V(\bm{\sigma}_{i})\label{eq:objective function}\\
\text{subject to} & \quad\quad\sigma_{i,m}^{2}=\max\{0,\mathbf{a}_{i}^{T}\mathbf{c}_{m}+b_{i}-\bar{G}_{i,m}^{2}\}\label{eq:constraint 1-1}\\
 & \quad\quad\sum_{m=1}^{M}\sigma_{i,m}^{2}\leq\mu\sum_{m=1}^{M}\bar{G}_{i,m}^{2}.\label{eq:constraint 3-1}
\end{align}
where $\rho$ is a non-negative hyper parameter that controls the
trade-off between the two competing privacy objectives, and (\ref{eq:constraint 3-1})
controls the total noise budget.

Intuitively, the slope $\mathbf{a}_{i}$ should be {\em aligned}
with the localization bias of the adversary under the noise-free gradient
$\mathbf{g}_{i}$. Consider the example illustrated in Fig. \ref{fig:Linear solution}(a),
where the location estimated by the adversary (\ref{eq:WCL}), or
a more general WCL strategy, has a positive bias. If the slope $\mathbf{a}_{i}$
is also positive, then the bias can be enlarged by adding noise to
the gradient. Such a property is formally captured by theorem in the
next subsection.

\subsection{Algorithm for Adaptive Noise Allocation}

The first term $P(\bm{\sigma}_{i})$ in the objective function (\ref{eq:objective function})
can be decomposed as follows.
\begin{thm}[Bias-Variance Decomposition]
\label{thm:localization error decomposition}As the number of grid
cells $M\to\infty$ with a decreasing grid size, the squared localization
error converges in probability to 
\begin{equation}
\|\tilde{\mathbf{p}}_{\text{u},i}-\mathbf{p}_{\text{u},i}\|_{2}^{2}\to\frac{1}{(1+\mu)^{2}}\|\Delta_{\text{g}}+\mu\Delta_{\text{n}}\|_{2}^{2}
\end{equation}
where the ``bias'' term is given as
\begin{equation}
\Delta_{\text{g}}=\frac{\sum_{m=1}^{M}\bar{G}_{i,m}^{2}\mathbf{c}_{m}}{\sum_{m=1}^{M}\bar{G}_{i,m}^{2}}-\mathbf{p}_{\text{u},i}
\end{equation}
and the ``variance'' term due to the artificial noise is
\begin{equation}
\Delta_{\text{n}}=\frac{\sum_{m=1}^{M}\sigma_{i,m}^{2}\mathbf{c}_{m}}{\sum_{m=1}^{M}\sigma_{i,m}^{2}}-\mathbf{p}_{\text{u},i}.
\end{equation}
\end{thm}
\begin{proof}
See Appendix \ref{sec:Proof-of-Theorem}.
\end{proof}
Theorem~\ref{thm:localization error decomposition} decomposes the
total localization error into a weighted combination of two distinct
vectors $\Delta_{\text{g}}$ and $\Delta_{\text{n}}$. The key insight
is that the magnitude of the resulting error vector $\Delta_{\text{g}}+\mu\Delta_{\text{n}}$
is asymptotically maximized when the ``variance'' vector $\Delta_{\text{n}}$
is {\em geometrically aligned} with the ``bias'' vector $\Delta_{\text{g}}$.
This principle provides an initial direction for the algorithm developed
next to solve for the optimal plane parameters.

Theorem~\ref{thm:localization error decomposition} provides an initialization
strategy for an iterative algorithm that solves (\ref{eq:objective function})
under finite $M$. Specifically, the slope $\mathbf{a}_{i}$ can be
initialized as $\mathbf{a}_{i}=r_{i}\mathbf{u}_{i}$, where $\mathbf{u}_{i}=\Delta_{\text{g}}/\|\Delta_{\text{g}}\|_{2}$.
The objective function (\ref{eq:objective function}) can be written
as $J(\mathbf{u}_{i},r_{i},b_{i})=P(\bm{\sigma}_{i}(\mathbf{u}_{i},r_{i},b_{i}))-\rho V(\bm{\sigma}_{i}(\mathbf{u}_{i},r_{i},b_{i}))$.
The algorithm can be developed using a nested structure, where in
the outer loop, the unit direction vector $\mathbf{u}_{i}$ is iteratively
updated using gradient descent. In the inner loop, for a fixed direction
$\mathbf{u}_{i}$, the corresponding scalar coefficient $r_{i}$ is
optimized. Furthermore, for any given pair $(\mathbf{u}_{i},r_{i})$,
the associated parameter $b_{i}$ is uniquely determined and can be
efficiently computed using a bisection search.

\begin{algorithm}
Input: Clipped local gradient $\bar{\mathbf{g}}_{i}$; user location
$\mathbf{p}_{\text{u},i}$; privacy parameter $\mu$; trade-off control
parameter $\rho$

Output: Differentially private gradient $\tilde{\mathbf{g}}_{i}$
\begin{enumerate}
\item Initialize the direction as $\ensuremath{\mathbf{u}_{i}\gets\frac{\Delta_{\text{g}}}{\|\Delta_{\text{g}}\|_{2}}\in\mathbb{R}^{2}}$
and the step as $r_{i}\gets r_{\max}$.
\item For the optimization of $\mathbf{u}_{i}$ in outer loop do
\begin{enumerate}
\item For the optimization of $r_{i}$ in inner loop do
\begin{enumerate}
\item Solve for $b_{i}$ given unique $(\mathbf{u}_{i},r_{i})$ in
\begin{equation}
{\scriptstyle \sum_{m=1}^{M}\max\{0,r_{i}\mathbf{u}_{i}^{\text{T}}\mathbf{c}_{m}+b_{i}-\bar{G}_{i,m}^{2}\}-\sum_{m=1}^{M}\mu\bar{G}_{i,m}^{2}=0}
\end{equation}
via bisection method
\item Compute the numerical gradient for $r_{i}$ as
\begin{equation}
\frac{\partial J}{\partial r_{i}}=\frac{1}{2\epsilon}(J(\mathbf{u}_{i},r_{i}+\epsilon)-J(\mathbf{u}_{i},r_{i}-\epsilon))
\end{equation}
\item Update $r_{i}$ using gradient ascent
\end{enumerate}
\item End For until $\|\partial J/\partial r_{i}\|_{2}\leq\tau_{r}$.
\item Compute the numerical gradient for $\mathbf{u}_{i}$ as
\begin{equation}
\nabla J=\left[\begin{array}{c}
\frac{1}{2\epsilon}(J(\mathbf{u}_{i}+\epsilon\mathbf{e}_{1},r_{i})-J(\mathbf{u}_{i}-\epsilon\mathbf{e}_{1},r_{i}))\\
\frac{1}{2\epsilon}(J(\mathbf{u}_{i}+\epsilon\mathbf{e}_{2},r_{i})-J(\mathbf{u}_{i}-\epsilon\mathbf{e}_{2},r_{i}))
\end{array}\right]
\end{equation}
where $\mathbf{e}_{1}=[1,0]^{\text{T}},\mathbf{e}_{2}=[0,1]^{\text{T}}$are
two orthogonal unit vectors in the 2D plane.
\item Update $\mathbf{u}_{i}$ using the gradient ascent
\end{enumerate}
\item End For until $\|\nabla J\|_{2}\leq\tau_{u}$.
\item The optimal parameters $(\mathbf{u}_{i}^{*},r_{i}^{*})$ are found,
find $b_{i}^{*}$ for $(\mathbf{u}_{i}^{*},r_{i}^{*})$ via bisection
search, and get the optimal noise allocation as
\begin{equation}
\sigma_{i,m}^{*2}=\max\{0,r_{i}^{*}\mathbf{u}_{i}^{*\text{T}}\mathbf{c}_{m}+b_{i}^{*}-\bar{G}_{i,m}^{2}\}
\end{equation}
\item Noise Generation:
\begin{enumerate}
\item Generate noise vector $\mathbf{n}_{i}\in\mathbb{R}^{M}$ where each
component $n_{i,m}\sim\mathcal{N}(0,\sigma_{i,m}^{*2})$.
\item Construct the final noisy gradient: $\tilde{\mathbf{g}}_{i}\gets\bar{\mathbf{g}}_{i}+\mathbf{n}_{i}$.
\end{enumerate}
\end{enumerate}
\caption{Geometry-Aligned Differential Privacy \label{alg:adaptive noise allocation}}
\end{algorithm}

The proposed geometry-aligned differential privacy algorithm is guaranteed
to converge to a stationary, locally optimal solution. At each iteration,
the algorithm uses a one-dimensional bisection search to solve a monotonic
root-finding problem for $b_{i}$, ensuring a unique solution for
this parameter. Building on this, gradient ascent updates for $r_{i}$
and $\mathbf{u}_{i}$, within a compact search space, monotonically
increase the continuous objective function $J(\mathbf{u}_{i},r_{i},b_{i})$.
Since $J$ is bounded above due to finite localization error and spatial
variance, the resulting non-decreasing sequence of objective values
will converge. Therefore, the algorithm converges to a stationary
point $(\mathbf{u}_{i}^{*},r_{i}^{*},b_{i}^{*})$ that satisfies the
first-order optimality conditions of the privacy maximization problem.

\section{Numerical Results}

In this section, we present simulation results to validate the performance
of our proposed geometry-aligned differential privacy algorithm on
radio map federated construction. The simulations are conducted in
a detailed 300 m $\times$ 300 m dense urban environment, populated
with buildings of diverse geometries, including circular, cubic, and
irregular shapes, and heights ranging from 10 m to 130 m. This virtual
city comprises 100 randomly distributed ground users and 200 aerial
base stations operating at a 50 m altitude. Each user collects 200
channel measurements from these base stations, and the entire area
is discretized into a grid of 3 m $\times$ 3 m cells for the purpose
of model reconstruction. The subsequent sections will analyze the
radio map accuracy and the effectiveness of our privacy-preserving
mechanism within this setting.

\subsection{Radio Map Construction Performance}

In this section, we evaluate the impact of the privacy-preserving
noise, governed by the parameter $\mu$ on the radio map reconstruction
accuracy. We compare its performance with the standard \ac{fedavg}
algorithm, which serves as a baseline and represents the ideal performance
without noise. In the evaluation, the noise allocation parameter is
fixed at $\rho=50$ and the clipping threshold is set to $C=1$. We
train the geometry-aligned differential privacy algorithm under a
range of distinct privacy budgets, by varying $\mu$ across the set
$\{0.5,1,5,10,20,50\}$. The reconstruction performance is quantified
by the \ac{mae} of the predicted channel gain over the entire geographical
area. The convergence behavior for different $\mu$ values is illustrated
by the \ac{mae} during each training iteration, with detailed results
presented in Table~\ref{tab:Radio Map Reconstruction MAE (dB) vs. Training Epoch for Different Privacy Budgets}.
Table~\ref{tab:Radio Map Reconstruction MAE (dB) vs. Training Epoch for Different Privacy Budgets}
shows that the non-private \ac{fedavg} baseline achieves a final
\ac{mae} of $3.81$ dB, while our geometry-aligned \ac{dp} method
reaches $4.25$ dB at $\mu=50$. This $0.44$ dB increase is marginal,
demonstrating a highly favorable privacy-utility trade-off.

\begin{table}
\caption{Radio Map Reconstruction \ac{mae} (dB) vs. Training Epoch for Different
Privacy Budgets $\mu$ \label{tab:Radio Map Reconstruction MAE (dB) vs. Training Epoch for Different Privacy Budgets}}

\centering{}\centerline{%
\begin{tabular}{cccccc}
\toprule 
 & $\mu$ & Epoch 1 & Epoch 50 & Epoch 100 & Epoch 200\tabularnewline
\midrule
\midrule 
\multirow{3}{*}{Proposed} & 1 & 13.68 & 6.72 & 4.01 & 3.92\tabularnewline
\cmidrule{2-6}
 & 10 & 13.68 & 6.79 & 4.11 & 4.05\tabularnewline
\cmidrule{2-6}
 & 50 & 13.68 & 6.85 & 4.21 & 4.25\tabularnewline
\midrule 
\ac{fedavg} &  & 13.68 & 6.68 & 3.98 & 3.81\tabularnewline
\bottomrule
\end{tabular}}
\end{table}

\subsection{Location Privacy Protection Performance}

This section evaluates the privacy protection afforded by our proposed
geometry-aligned differential privacy algorithm against the \ac{wcl}
adversary described in Section 3.2. The performance is measured by
the estimated localization error by adversary\textemdash the distance
between a user's true location and the one inferred by the \ac{wcl}
attack. A greater error corresponds to a higher degree of privacy.
We analyze the influence of two key parameters: the total noise budget
$\mu$ and the geometric trade-off parameter $\rho$, which controls
the spatial allocation of noise. Throughout the evaluation, the clipping
threshold is fixed at $C=1$.

\subsubsection{Impact of Noise Budget $\mu$ on Privacy}

\begin{figure}
\centerline{\includegraphics[width=0.95\columnwidth]{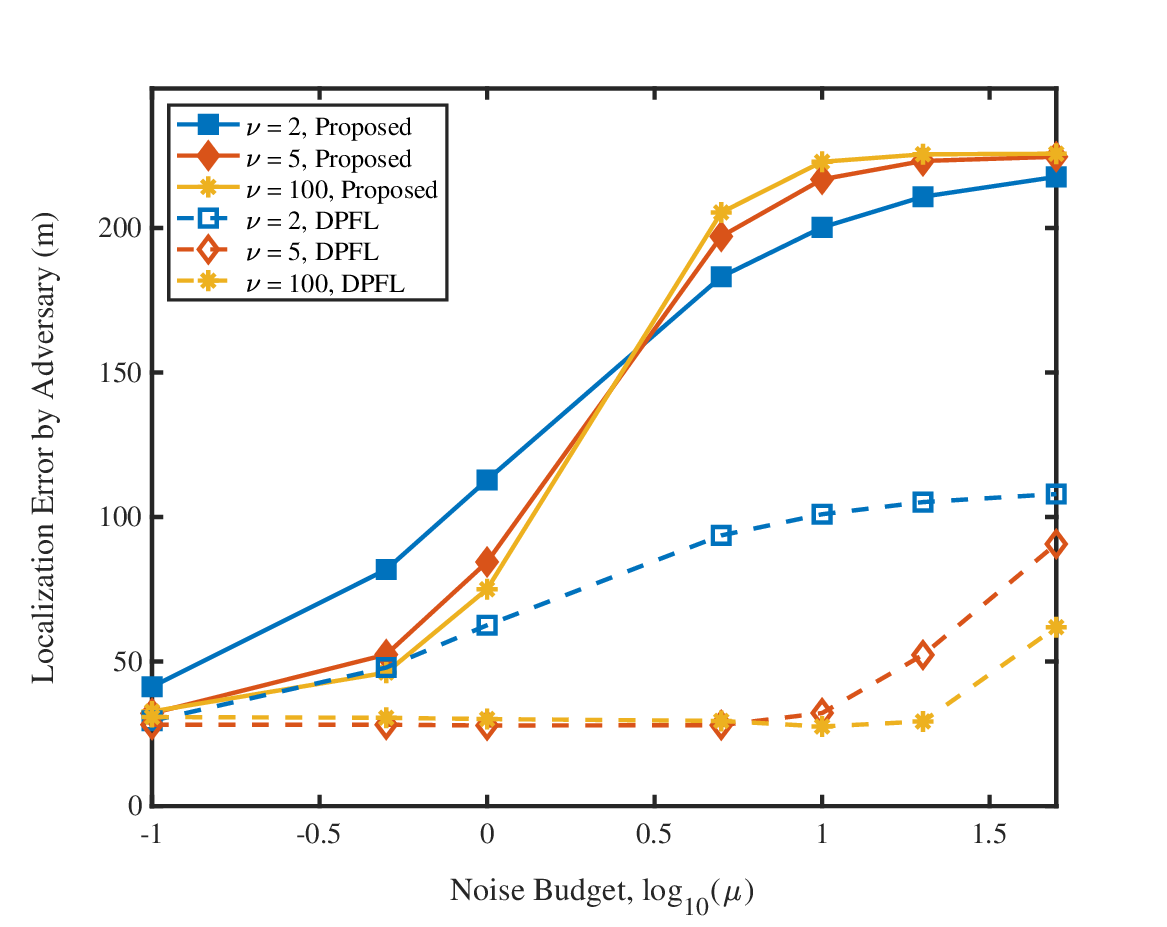}}\caption{Localization error by adversary vs. noise budget $\mu$ for different
parameter $\nu$ used by the adversary. The proposed scheme achieves
substantially better protection over all parameters.}

\label{fig:Estimated Localization Error by Adversary vs. Noise Budget =00005Cmu  for Different =00005Cac=00007Bwcl=00007D Power Parameter =00005Cnu }
\end{figure}
We evaluate the impact of the noise budget $\mu$ by comparing our
geometry-aligned differential privacy algorithm with a uniform-noise
\ac{dpfl} baseline. In this analysis, the geometric trade-off parameter
of the proposed algorithm is fixed to $\rho=1$. As shown in Fig.~\ref{fig:Estimated Localization Error by Adversary vs. Noise Budget =00005Cmu  for Different =00005Cac=00007Bwcl=00007D Power Parameter =00005Cnu },
increasing the noise budget $\mu$ raises the adversary's localization
error for both the geometry-aligned \ac{dp} scheme and the uniform
\ac{dpfl} baseline. For any given $\mu$, the error induced by our
strategy is substantially larger than that from \ac{dpfl} across
all tested attacker power values $\nu$.

\begin{figure}
\centerline{\includegraphics[width=0.95\columnwidth]{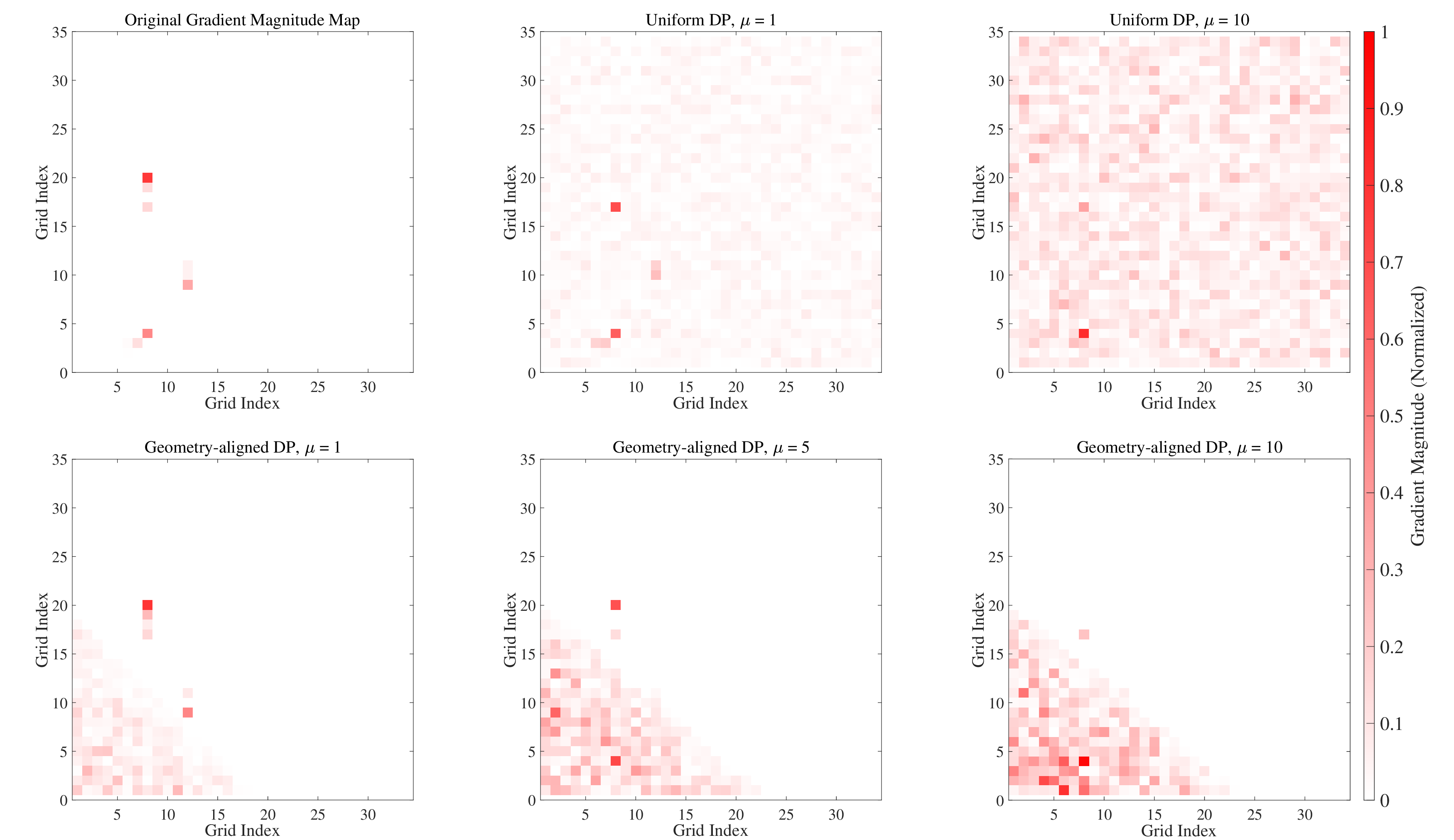}}\caption{Gradient magnitude maps under increasing noise budgets for the uniform
\ac{dp} baseline (top row) and the geometry-aligned \ac{dp} scheme
(bottom row). As $\mu$ increases, the geometry-aligned \ac{dp} obfuscates
spatial gradient patterns while controlling the adversary\textquoteright s
localization error, illustrating the trade-off between pattern concealment
and the numerical location-privacy metric.}

\label{fig:Visual comparison of spatial gradient obfuscation between standard =00005Cac=00007Bdpfl=00007D (top row) and =00005Cac=00007Bg-adpfl=00007D (bottom row). }
\end{figure}
This robust privacy gain stems from injecting structured noise that
shifts the gradient\textquoteright s perceived centroid rather than
simply diffusing it. In contrast to uniform noise, which merely obscures
the original gradient, the geometry-aligned noise produces a deliberate
spatial pattern that becomes more pronounced with larger $\mu$ as
shown in Fig.~\ref{fig:Visual comparison of spatial gradient obfuscation between standard =00005Cac=00007Bdpfl=00007D (top row) and =00005Cac=00007Bg-adpfl=00007D (bottom row). }.
This pattern actively misleads the \ac{wcl} adversary and amplifies
the localization error.

\subsubsection{Efficacy of Adaptive Noise Allocation $\rho$}

\begin{figure}
\centerline{\includegraphics[width=0.95\columnwidth]{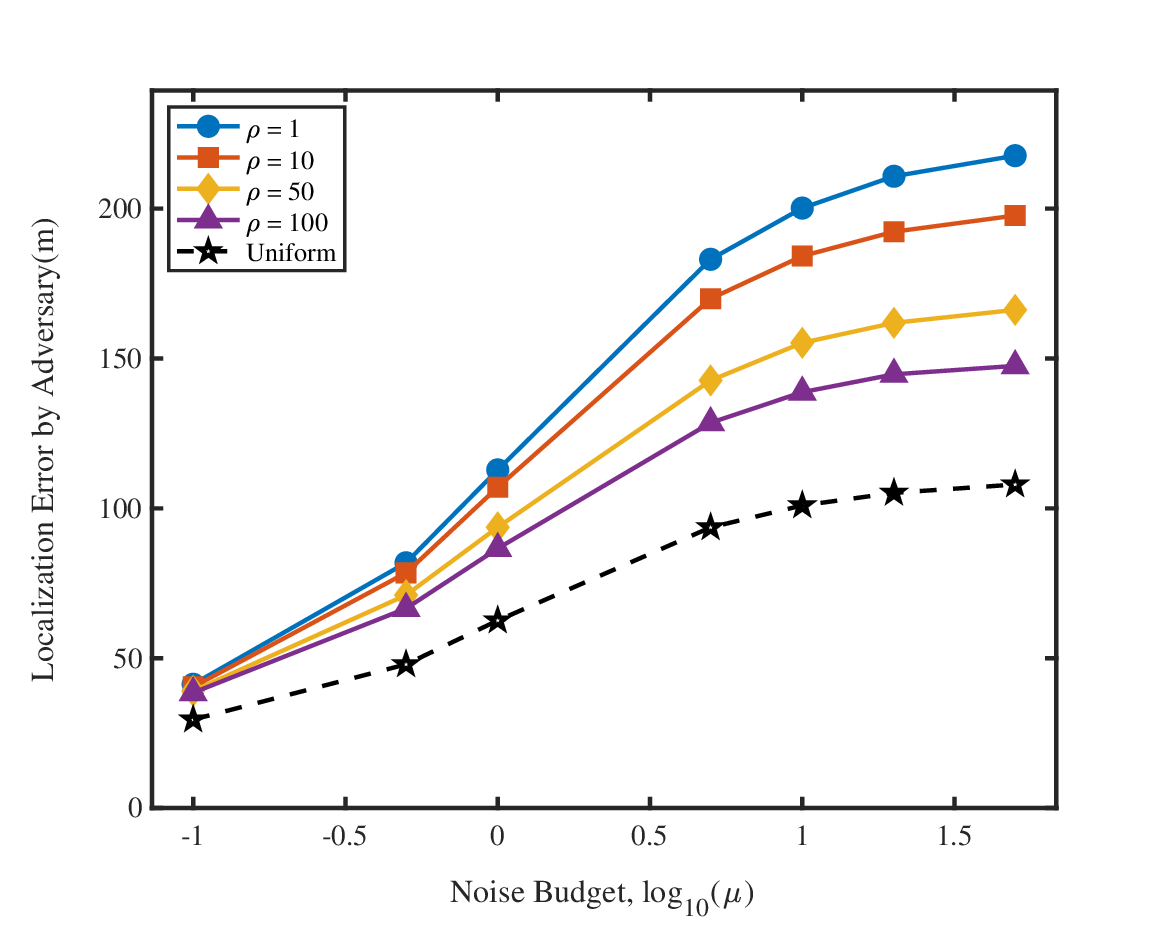}}\caption{\label{fig:Impact of the geometric trade-off parameter =00005Crho  on the estimated localization error by adversary.}Estimated
adversarial localization error vs. geometric trade-off parameter $\rho$
(with adversary power $\nu=2$). Smaller $\rho$ values produce the
higher localization errors, whereas larger \textgreek{\textrho} slightly
reduces the error; all proposed noise allocations still outperform
the uniform noise baseline in terms of adversary error.}
\end{figure}
Fig.~\ref{fig:Impact of the geometric trade-off parameter =00005Crho  on the estimated localization error by adversary.}
shows that smaller values of the trade-off parameter $\rho$ produce
higher adversarial localization errors. With $\rho=1$, the noise
is concentrated in a high-impact region, yielding the maximum adversary
error exceeding $220$ m at large $\mu$. Increasing $\rho$ spreads
the noise more evenly, which slightly reduces the peak error to approximately
$150$m at $\rho=100$ but results in a more diffuse gradient perturbation.
This effect is evident in the gradient maps: a small $\rho$ creates
a localized noise pattern, whereas a large $\rho$ distributes noise
broadly. Importantly, every geometry-aligned configuration still induces
much higher adversary error than the uniform baseline.

\section{Conclusion}

This paper investigated the vulnerability of federated radio map construction
to location inference attacks. Spatially decaying gradient patterns
were identified as a key source of location information leakage. To
counter this threat, we proposed a geometry-aligned differential privacy
framework. This method injects heterogeneous noise into the model
updates to obscure spatial gradient patterns and degrade the localization
performance of adversary strategies, thus enhancing the location privacy
of the users. A theoretical convergence bound was derived to quantify
the trade-off between privacy and utility for the proposed approach.
It is shown that the federated training algorithm converges to a stationary
point with a bounded expected gradient norm under the proposed non-uniform
noise scheme. Simulations showed that the framework can improve location
privacy by raising the adversary\textquoteright s localization error
from $30$m to over $180$m. This privacy gain was achieved with only
a $0.2$dB degradation in map accuracy compared to a uniform noise
baseline.

\appendices{}

\section{Proof of Theorem \ref{thm:gradient attenuation}\label{sec:Proof-of-Theorem1}}

Denote the component of the local gradient element as
\begin{equation}
g_{i,j,m}=\Gamma(\mathbf{p}_{j};\mathbf{h})D(\mathbf{p}_{j};\mathbf{h},m).\label{eq:local gradient component}
\end{equation}
Thus, the local gradient element (\ref{eq:simplification of local gradient element})
simplifies to
\begin{equation}
G_{i,m}=\frac{2}{J_{i}}\sum_{j\in\mathcal{N}_{i,m}}g_{i,j,m}.\label{eq:local gradient expansion}
\end{equation}

To facilitate the monotonicity discussion of $G_{i,m}$, we first
investigate the sign property of the component $g_{i,j,m}$. Based
on (\ref{eq:g_i,m first term}), according that the actual channel
gain is formulated as $y_{j}=\gamma(\mathbf{p}_{j},\mathbf{h}^{*})+\xi_{j}$,
the product term $\Gamma(\mathbf{p}_{j};\mathbf{h})$ in the component
$g_{i,j,m}$ is split into
\begin{equation}
\begin{aligned}\Gamma(\mathbf{p}_{j};\mathbf{h})=\hat{\Gamma}(\mathbf{p}_{j};\mathbf{h})+\hat{\xi}_{j}\end{aligned}
\label{eq:first term}
\end{equation}
where
\begin{equation}
\begin{aligned}\hat{\Gamma}(\mathbf{p}_{j};\mathbf{h}) & =(\gamma(\mathbf{p}_{j};\mathbf{h}^{*})-\gamma(\mathbf{p}_{j};\mathbf{h}))\\
 & \ \ \ \times((\beta_{1}+\alpha_{1}\log_{10}d(\mathbf{p}_{j}))\\
 & \ \ \ \ \ -(\beta_{0}+\alpha_{0}\log_{10}d(\mathbf{p}_{j})))\\
 & =((\beta_{1}+\alpha_{1}\log_{10}d(\mathbf{p}_{j}))-(\beta_{0}+\alpha_{0}\log_{10}d(\mathbf{p}_{j})))^{2}\\
 & \ \ \ \times(S(\mathbf{p}_{j};\mathbf{h}^{*})-S(\mathbf{p}_{j};\mathbf{h}))
\end{aligned}
\end{equation}
is the model error term whose sign depends on the model difference
$(S(\mathbf{p}_{j};\mathbf{h}^{*})-S(\mathbf{p}_{j};\mathbf{h}))$,
and
\begin{equation}
\hat{\xi}_{j}=\xi_{j}((\beta_{1}+\alpha_{1}\log_{10}d(\mathbf{p}_{j}))-(\beta_{0}+\alpha_{0}\log_{10}d(\mathbf{p}_{j})))
\end{equation}
is the noise term, a product of Gaussian noise and a constant, still
Gaussian noise. Thus, based on (\ref{eq:local gradient component}),
(\ref{eq:first term}), the component of the local gradient element
$g_{i,j,m}$ is split into the real model value and the noise as
\begin{equation}
\begin{aligned}g_{i,j,m} & =\tilde{g}_{i,j,m}+\tilde{\xi}_{i,j,m}\\
 & =\hat{\Gamma}(\mathbf{p}_{j};\mathbf{h})D(\mathbf{p}_{j};\mathbf{h},m)+\hat{\xi}_{j}D(\mathbf{p}_{j};\mathbf{h},m)
\end{aligned}
\label{eq:noisy term}
\end{equation}
where the noise term $\tilde{\xi}_{i,j,m}$ is the product of Gaussian
noise and a constant, still Gaussian noise. According to (\ref{eq:g_i,m second term}),
the term $D(\mathbf{p}_{j};\mathbf{h},m)$ is always positive.

Therefore, for the expansion of the model error term $\tilde{g}_{i,j,m},\forall j\in\mathcal{N}_{i,m}$
in (\ref{eq:noisy term}), the product terms except for $(S(\mathbf{p}_{j};\mathbf{h}^{*})-S(\mathbf{p}_{j};\mathbf{h}))$
are always positive. To explore the sign of $g_{i,j,m}$, we only
focus on the sign of the difference term $(S(\mathbf{p}_{j};\mathbf{h}^{*})-S(\mathbf{p}_{j};\mathbf{h}))$.
According to (\ref{eq:g_i,m second term}), (\ref{eq:virtual-obsatcle}),
when $\mathbf{h}^{*}\prec\mathbf{h}$, the expectations of the contributing
gradient components $\tilde{g}_{i,j,m},\forall j\in\mathcal{N}_{i,m}$
are consistently greater than $0$. On the contrary, when $\mathbf{h}^{*}\succ\mathbf{h}$,
the expectations of the contributing gradient components $\tilde{g}_{i,j,m},\forall j\in\mathcal{N}_{i,m}$
are consistently less than $0$.

Next we analyze the property of the sum $G_{i,m}$. According to the
expansion (\ref{eq:local gradient expansion}), the expectation of
the $m$th local gradient element is formulated as
\begin{equation}
\mathbb{E}\{G_{i,m}^{2}\}=\Big(\frac{2}{J_{i}}\Big)^{2}\mathbb{E}\Big\{\Big(\sum_{j\in\mathcal{N}_{i,m}}g_{i,j,m}\Big)^{2}\Big\}.
\end{equation}
Combined with (\ref{eq:noisy term}), it is further computed as
\begin{align*}
\mathbb{E}\{G_{i,m}^{2}\} & =\Big(\frac{2}{J_{i}}\Big)^{2}\mathbb{E}\Big\{\Big(\sum_{j\in\mathcal{N}_{i,m}}(\tilde{g}_{i,j,m}+\tilde{\xi}_{i,j,m})\Big)^{2}\Big\}.
\end{align*}
Thus, the expectation of the local gradient element $\mathbb{E}\{G_{i,m}^{2}\}$
is extended into
\begin{align*}
\mathbb{E}\{G_{i,m}^{2}\} & =\Big(\frac{2}{J_{i}}\Big)^{2}\Big(\mathbb{E}\Big\{\sum_{j\in\mathcal{N}_{i,m}}\tilde{g}_{i,j,m}^{2}\Big\}\\
 & \ \ +\mathbb{E}\Big\{\sum_{j\in\mathcal{N}_{i,m}}\tilde{\xi}_{i,j,m}^{2}\Big\}\\
 & \ \ +\mathbb{E}\bigg\{\sum_{\substack{j,j'\in\mathcal{N}_{i,m}\\
j\ne j'
}
}\tilde{g}_{i,j,m}\tilde{g}_{i,j',m}\bigg\}\Big).
\end{align*}
The expectation of the ($m-1)$th element of the local gradient is
computed as
\begin{align*}
\mathbb{E}\{G_{i,m-1}^{2}\} & =\Big(\frac{2}{J_{i}}\Big)^{2}\Big(\mathbb{E}\Big\{\sum_{j\in\mathcal{N}_{i,m-1}}\hat{g}_{i,j,m-1}^{2}\Big\}\\
 & \ \ +\mathbb{E}\Big\{\sum_{j\in\mathcal{N}_{i,m-1}}\hat{\xi}_{j}^{2}\Big\}\\
 & \ \ +\mathbb{E}\bigg\{\sum_{\substack{j,j'\in\mathcal{N}_{i,m-1}\\
j\ne j'
}
}\hat{g}_{i,j,m-1}\hat{g}_{i,j',m-1}\bigg\}\Big).
\end{align*}

As established previously, all $\tilde{g}_{i,j,m}$ for $j\in\mathcal{N}_{i,m}$
share the same sign. Consequently, this cross-product term is non-negative.
We now compare the terms in the expansions of $\mathbb{E}\{G_{i,m}^{2}\}$
and $\mathbb{E}\{G_{i,m-1}^{2}\}$. The geometric property of spatial
coverage implies a nesting relationship for the sets of influential
measurements 
\begin{equation}
\mathcal{N}_{i,m}\subseteq\mathcal{N}_{i,m-1}.
\end{equation}
Since all three components in the expansion of $\mathbb{E}\{G_{i,m}^{2}\}$
are composed of non-negative summation terms, the reduction in the
size of the summation set from $\mathcal{N}_{i,m-1}$ to $\mathcal{N}_{i,m}$
directly implies: 
\begin{equation}
\mathbb{E}\{G_{i,m}^{2}\}\leq\mathbb{E}\{G_{i,m-1}^{2}\}.
\end{equation}
This completes the proof.

\section{Proof of Lemma \ref{lemma:loss_reduction_bound}\label{sec:Proof-of-Lemma1}}

We first derive the expectation of the decrease in the loss function
with respect to $\mathbf{h}(t)$ given $\boldsymbol{\theta}(t)$ in
each round of training. First, by using the $L$-smoothness property
in (\ref{eq:L-smooth}), we have
\begin{equation}
\begin{aligned}\mathbb{E}\{F(\mathbf{h}(t+1),\boldsymbol{\theta}(t))\}\quad\quad\quad\quad\quad\quad\quad\quad\quad\quad\quad\quad\quad\quad\\
\le\mathbb{E}\{F(\mathbf{h}(t),\boldsymbol{\theta}(t))\}-\mathbb{E}\{\langle\mathbf{g}(t),\mathbf{h}(t+1)-\mathbf{h}(t)\rangle\}\\
+\frac{L}{2}\mathbb{E}\{\|\mathbf{h}(t+1)-\mathbf{h}(t)\|_{2}^{2}\}.
\end{aligned}
\label{eq:lemma1_smooth}
\end{equation}
Recall that the update local gradient is $\tilde{\mathbf{g}}_{i}(t)=\bar{\mathbf{g}}_{i}(t)+\mathbf{n}_{i}(t)$.
Under the DP policy, the aggregation for $\mathbf{h}(t+1)$ in (\ref{eq:h aggregation})
becomes
\begin{equation}
\begin{aligned}\mathbf{h}(t+1) & =\mathbf{h}(t)+\eta_{h}\sum_{i=1}^{N}\frac{J_{i}}{J}\tilde{\mathbf{g}}_{i}(t)\\
 & =\mathbf{h}(t)+\eta_{h}\sum_{i=1}^{N}\frac{J_{i}}{J}\bar{\mathbf{g}}_{i}(t)+\eta_{h}\sum_{i=1}^{N}\frac{J_{i}}{J}\mathbf{n}_{i}(t).
\end{aligned}
\end{equation}
 In this case, (\ref{eq:lemma1_smooth}) is derived as
\begin{equation}
\begin{aligned}\mathbb{E}\{F(\mathbf{h}(t+1),\boldsymbol{\theta}(t))\}\quad\quad\quad\quad\quad\quad\quad\quad\quad\quad\quad\quad\\
\le\mathbb{E}\{F(\mathbf{h}(t),\boldsymbol{\theta}(t))\}\quad\quad\quad\quad\quad\quad\quad\quad\quad\quad\quad\quad\\
-\eta_{h}\mathbb{E}\{\langle\mathbf{g}(t),\sum_{i=1}^{N}\frac{J_{i}}{J}\bar{\mathbf{g}}_{i}(t)+\eta_{h}\sum_{i=1}^{N}\frac{J_{i}}{J}\mathbf{n}_{i}(t)\rangle\}\\
+\frac{L\eta_{h}^{2}}{2}\mathbb{E}\{\|\sum_{i=1}^{N}\frac{J_{i}}{J}\bar{\mathbf{g}}_{i}(t)+\sum_{i=1}^{N}\frac{J_{i}}{J}\mathbf{n}_{i}(t)\|_{2}^{2}\}.
\end{aligned}
\label{eq:L-smooth-variation}
\end{equation}
Since the expectation of the noise $\mathbb{E}\{\mathbf{n}_{i}(t)\}=0$
and the noise is independent of the gradients, (\ref{eq:L-smooth-variation})
is simplified as
\begin{equation}
\begin{aligned}\mathbb{E}\{F(\mathbf{h}(t+1),\boldsymbol{\theta}(t))\}\quad\quad\quad\quad\quad\quad\quad\quad\quad\quad\quad\quad\quad\quad\quad\\
\le\mathbb{E}\{F(\mathbf{h}(t),\boldsymbol{\theta}(t))\}-\eta_{h}\mathbb{E}\{\langle\mathbf{g}(t),\sum_{i=1}^{N}\frac{J_{i}}{J}\bar{\mathbf{g}}_{i}(t)\rangle\}\quad\quad\\
+\frac{L\eta_{h}^{2}}{2}(\mathbb{E}\{\|\sum_{i=1}^{N}\frac{J_{i}}{J}\bar{\mathbf{g}}_{i}(t)\|_{2}^{2}\}+\mathbb{E}\{\|\sum_{i=1}^{N}\frac{J_{i}}{J}\mathbf{n}_{i}(t)\|_{2}^{2}\}).
\end{aligned}
\label{eq:L-smooth-simplification}
\end{equation}

Now, we first simplify the last line of (\ref{eq:L-smooth-simplification}).
For the first term $\mathbb{E}\{\|\sum_{i=1}^{N}\frac{J_{i}}{J}\bar{\mathbf{g}}_{i}(t)\|_{2}^{2}\}$,
we use Jensen's inequality and the fact that clipping only reduces
the norm, i.e., $\|\bar{\mathbf{g}}_{i}(t)\|_{2}\le\|\mathbf{g}_{i}(t)\|_{2}$,
and derive that
\begin{equation}
\begin{aligned}\mathbb{E}\{\|\sum_{i=1}^{N}\frac{J_{i}}{J}\bar{\mathbf{g}}_{i}(t)\|_{2}^{2}\} & \le\mathbb{E}\{\sum_{i=1}^{N}\frac{J_{i}}{J}\|\bar{\mathbf{g}}_{i}(t)\|_{2}^{2}\}\\
 & \le\sum_{i=1}^{N}\frac{J_{i}}{J}\mathbb{E}\{\|\mathbf{g}_{i}(t)\|_{2}^{2}\}.
\end{aligned}
\label{eq:lemma1_signal}
\end{equation}
For the second term, using the independence of noise across users
and Assumption 2
\begin{equation}
\begin{aligned}\mathbb{E}\{\|\sum_{i=1}^{N}\frac{J_{i}}{J}\mathbf{n}_{i}(t)\|_{2}^{2}\} & =\sum_{i=1}^{N}\frac{J_{i}^{2}}{J^{2}}\mathbb{E}\{\|\mathbf{n}_{i}(t)\|_{2}^{2}\}\\
 & \le\sum_{i=1}^{N}\frac{J_{i}^{2}}{J^{2}}\mu\mathbb{E}\{\|\bar{\mathbf{g}}_{i}(t)\|_{2}^{2}\}.
\end{aligned}
\label{eq:lemma1_noise}
\end{equation}
Combining (\ref{eq:lemma1_signal}) and (\ref{eq:lemma1_noise}) back
into (\ref{eq:L-smooth-simplification}), we have
\begin{equation}
\begin{aligned}\mathbb{E}\{F(\mathbf{h}(t+1),\boldsymbol{\theta}(t))\}\quad\quad\quad\quad\quad\quad\quad\quad\quad\quad\quad\quad\quad\quad\quad\\
\le\mathbb{E}\{F(\mathbf{h}(t),\boldsymbol{\theta}(t))\}-\eta_{h}\mathbb{E}\{\langle\mathbf{g}(t),\sum_{i=1}^{N}\frac{J_{i}}{J}\bar{\mathbf{g}}_{i}(t)\rangle\}\quad\quad\\
+\frac{L\eta_{h}^{2}}{2}\sum_{i=1}^{N}(\frac{J_{i}}{J}+\mu\frac{J_{i}^{2}}{J^{2}})\mathbb{E}\{\|\mathbf{g}_{i}(t)\|_{2}^{2}\}.
\end{aligned}
\end{equation}
Now, applying Assumption 1, we have
\begin{equation}
\begin{aligned}\mathbb{E}\{F(\mathbf{h}(t+1),\boldsymbol{\theta}(t))\}\quad\quad\quad\quad\quad\quad\quad\quad\quad\quad\quad\quad\quad\quad\quad\\
\le\mathbb{E}\{F(\mathbf{h}(t),\boldsymbol{\theta}(t))\}-\eta_{h}\mathbb{E}\{\langle\mathbf{g}(t),\sum_{i=1}^{N}\frac{J_{i}}{J}\bar{\mathbf{g}}_{i}(t)\rangle\}\quad\quad\\
+\frac{L\eta_{h}^{2}B^{2}}{2}(1+\mu\tilde{J})\mathbb{E}\{\|\mathbf{g}(t)\|_{2}^{2}\}.
\end{aligned}
\label{eq:L-smooth-simplification2}
\end{equation}
where we use $\sum\frac{J_{i}}{J}=1$ and $\tilde{J}=\sum\frac{J_{i}^{2}}{J^{2}}$.

We next simplify the cross-term in the second line of (\ref{eq:L-smooth-simplification2})
as
\begin{equation}
\mathbb{E}\langle\mathbf{g}(t),\sum_{i=1}^{N}\frac{J_{i}}{J}\bar{\mathbf{g}}_{i}(t)\rangle=\mathbb{E}\langle\mathbf{g}(t)\text{,\ensuremath{\bar{\mathbf{g}}}(t)}\rangle
\end{equation}
where $\bar{\mathbf{g}}(t)=\sum_{i=1}^{N}\frac{J_{i}}{J}\bar{\mathbf{g}}_{i}(t)$
denotes the aggregation of the clipped gradient.

To investigate the bias introduced by gradient clipping, we define
$\mathbf{b}_{i}(t)=\mathbf{g}_{i}(t)-\bar{\mathbf{g}}_{i}(t)$ be
the clipped-off portion for user $i$. Then, the global bias $\mathbf{b}(t)=\sum_{i=1}^{N}\frac{J_{i}}{J}\mathbf{b}_{i}(t)$.
If $\|\mathbf{g}_{i}(t)\|_{2}\leq C$, then $\ensuremath{\mathbf{b}_{i}(t)=\mathbf{0}}$.
If $\|\mathbf{g}_{i}(t)\|_{2}>C$, then $\mathbf{b}_{i}(t)=\mathbf{g}_{i}(t)\left(1-\frac{C}{\|\mathbf{g}_{i}(t)\|_{2}}\right).$
The norm is $\|\mathbf{b}_{i}(t)\|_{2}=\|\mathbf{g}_{i}(t)\|_{2}-C$,
which is bounded as
\begin{equation}
\|\mathbf{b}_{i}(t)\|_{2}=\frac{\|\mathbf{g}_{i}(t)\|_{2}^{2}-C^{2}}{\|\mathbf{g}_{i}(t)\|_{2}+C}<\frac{\|\mathbf{g}_{i}(t)\|_{2}^{2}}{2C}.
\end{equation}
This bound holds trivially when $\|\mathbf{g}_{i}(t)\|_{2}\le C$
as well. Using Jensen's inequality for the aggregated bias, the expectation
of the global bias norm is derived as
\begin{equation}
\begin{aligned}\mathbb{E}\{\|\mathbf{b}(t)\|_{2}^{2}\} & =\mathbb{E}\{\|\sum_{i=1}^{N}\frac{J_{i}}{J}\mathbf{b}_{i}(t)\|_{2}^{2}\}\le\mathbb{E}\{\sum_{i=1}^{N}\frac{J_{i}}{J}\|\mathbf{b}_{i}(t)\|_{2}^{2}\}\\
 & <\mathbb{E}\{\sum_{i=1}^{N}\frac{J_{i}}{J}(\frac{\|\mathbf{g}_{i}(t)\|_{2}^{2}}{2C})^{2}\}=\frac{1}{4C^{2}}\mathbb{E}\{\|\mathbf{g}_{i}(t)\|_{2}^{4}\}\\
 & \leq\frac{B^{4}L^{4}}{4C^{2}}.
\end{aligned}
\end{equation}
Thus, the cross-term in the second line $\mathbb{E}\langle\mathbf{g}(t)\text{,\ensuremath{\bar{\mathbf{g}}}(t)}\rangle$
is derived as
\begin{equation}
\begin{aligned}\mathbb{E}\{\langle\mathbf{g}(t),\bar{\mathbf{g}}(t)\rangle\} & =\mathbb{E}\{\langle\mathbf{g}(t),\mathbf{g}(t)-\mathbf{b}(t)\rangle\}\\
 & \ge\mathbb{E}\{\|\mathbf{g}(t)\|_{2}^{2}\}\\
 & \quad\quad\quad-\frac{1}{2}\left(\mathbb{E}\{\|\mathbf{g}(t)\|_{2}^{2}\}+\mathbb{E}\{\|\mathbf{b}(t)\|_{2}^{2}\}\right)\\
 & =\frac{1}{2}\mathbb{E}\{\|\mathbf{g}(t)\|_{2}^{2}\}-\frac{1}{2}\mathbb{E}\{\|\mathbf{b}(t)\|_{2}^{2}\}.
\end{aligned}
\label{eq:cross-term simplification}
\end{equation}
Substituting (\ref{eq:cross-term simplification}) into (\ref{eq:L-smooth-simplification2}),
we get
\begin{equation}
\begin{aligned}\mathbb{E}\{F(\mathbf{h}(t+1),\boldsymbol{\theta}(t))\}-\mathbb{E}\{F(\mathbf{h}(t),\boldsymbol{\theta}(t))\}\quad\quad\quad\quad\quad\quad\quad\\
\le-\eta_{h}(\frac{1}{2}\mathbb{E}\{\|\mathbf{g}(t)\|_{2}^{2}\}-\frac{1}{2}\mathbb{E}\{\|\mathbf{b}(t)\|_{2}^{2}\})\quad\quad\quad\quad\quad\quad\\
+\frac{L\eta_{h}^{2}}{2}B^{2}(1+\mu\tilde{J})\mathbb{E}\{\|\mathbf{g}(t)\|_{2}^{2}\}\quad\quad\quad\quad\quad\quad\quad\quad\quad\\
=-\eta_{h}(\frac{1}{2}-\frac{L\eta_{h}}{2}B^{2}(1+\mu\tilde{J}))\mathbb{E}\{\|\mathbf{g}(t)\|_{2}^{2}\}+\frac{\eta_{h}B^{4}L^{4}}{8C^{2}}.
\end{aligned}
\end{equation}
The proof of the first conclusion in Lemma \ref{lemma:loss_reduction_bound}
is completed.

Then, we analyze the update of $\boldsymbol{\theta}$ from $(\mathbf{h}(t+1),\boldsymbol{\theta}(t))$
to $(\mathbf{h}(t+1),\boldsymbol{\theta}(t+1))$. Using $L$-smoothness
with respect to $\boldsymbol{\theta}$, we have
\begin{equation}
\begin{aligned}F(\mathbf{h}(t+1),\boldsymbol{\theta}(t+1)) & \leq F(\mathbf{h}(t+1),\boldsymbol{\theta}(t))\\
 & \quad+\nabla F(\boldsymbol{\theta}(t);\mathbf{h}(t+1))^{\top}\\
 & \quad\quad\ \times(\boldsymbol{\theta}(t+1)-\boldsymbol{\theta}(t))\\
 & \quad+\frac{L}{2}||\boldsymbol{\theta}(t+1)-\boldsymbol{\theta}(t)||^{2}.
\end{aligned}
\label{eq:theta update}
\end{equation}
Recall the standard update rule for propagation parameter $\boldsymbol{\theta}$
in (\ref{eq:theta aggregation}), we get
\begin{equation}
\boldsymbol{\theta}(t+1)-\boldsymbol{\theta}(t)=-\eta_{\theta}\nabla F(\boldsymbol{\theta}(t);\mathbf{h}(t+1)).\label{eq:theta gradient}
\end{equation}
Substituting (\ref{eq:theta gradient}) into (\ref{eq:theta update}),
we have
\begin{equation}
\begin{aligned}F(\mathbf{h}(t+1),\boldsymbol{\theta}(t+1)) & \leq F(\mathbf{h}(t+1),\boldsymbol{\theta}(t))\\
 & \quad-\eta_{\theta}\|\nabla F(\boldsymbol{\theta}(t);\mathbf{h}(t+1))\|^{2}\\
 & \quad+\frac{L\eta_{\theta}^{2}}{2}\|\nabla F(\boldsymbol{\theta}(t);\mathbf{h}(t+1))\|^{2}.
\end{aligned}
\end{equation}
We take the expectation of the randomness introduced by the update
of $\mathbf{h}$, and get
\begin{equation}
\begin{aligned}\mathbb{E}\{F(\mathbf{h}(t+1),\boldsymbol{\theta}(t+1))-F(\mathbf{h}(t+1),\boldsymbol{\theta}(t))\}\\
\leq-\kappa_{\theta}\mathbb{E}\{\|\nabla F(\boldsymbol{\theta}(t);\mathbf{h}(t+1)\|^{2}\}
\end{aligned}
\label{eq:=00005Cthetaupdate}
\end{equation}
where $\kappa_{\theta}=\eta_{\theta}-\frac{L\eta_{\theta}^{2}}{2}$.

\section{Proof of Corollary \ref{cor:Convergence}\label{sec:Proof-of-Corollary}}

Recall the initial bound (\ref{eq:initial bound})
\begin{equation}
\begin{aligned}\frac{1}{T}\sum_{t=0}^{T-1}\mathbb{E}\{\|\nabla F(\mathbf{h}(t),\boldsymbol{\theta}(t))\|_{2}^{2}\}\quad\quad\quad\quad\quad\quad\quad\quad\quad\quad\quad\\
\leq\frac{\mathbb{E}\{F(\mathbf{h}(0),\boldsymbol{\theta}(0))-F(\mathbf{h}(T),\boldsymbol{\theta}(T))\}}{T\kappa}+\frac{\eta_{h}B^{4}L^{4}}{8C^{2}\kappa}.
\end{aligned}
\label{eq:initial bound-1}
\end{equation}
Since the difference of the function $F(\mathbf{h},\text{\ensuremath{\boldsymbol{\theta}}})$
is bounded, when $T\to\infty$, the first on the right hand side tends
to $0$. It is derived that
\begin{equation}
\frac{1}{T}\sum_{t=0}^{T-1}\mathbb{E}\{\|\nabla F(\mathbf{h}(t),\boldsymbol{\theta}(t))\|_{2}^{2}\}\leq\frac{\eta_{h}B^{4}L^{4}}{8C^{2}\kappa}
\end{equation}
Thus after a sufficient number of iterations, this bound is reduced
to
\begin{equation}
\mathbb{E}\{\|\nabla F(\mathbf{h}(t),\boldsymbol{\theta}(t))\|_{2}^{2}\}\leq\frac{\eta_{h}B^{4}L^{4}}{8C^{2}\kappa}.\label{eq:limit bound}
\end{equation}
Applying the Jensen's inequality, we have
\begin{equation}
\mathbb{E}\{\|\nabla F(\mathbf{h}(t),\boldsymbol{\theta}(t))\|_{2}\}^{2}\leq\mathbb{E}\{\|\nabla F(\mathbf{h}(t),\boldsymbol{\theta}(t))\|_{2}^{2}\}.\label{eq:expetation inequality}
\end{equation}
Combining (\ref{eq:limit bound}) and (\ref{eq:expetation inequality}),
we derive
\begin{equation}
\mathbb{E}\{\|\nabla F(\mathbf{h}(t),\boldsymbol{\theta}(t))\|\}\leq\sqrt{\frac{\eta_{h}B^{4}L^{4}}{8C^{2}\kappa}}.
\end{equation}
The proof is completed.

\section{Proof of Theorem \ref{thm:localization error decomposition}\label{sec:Proof-of-Theorem}}

We explore the approximation of the localization error by adversary
under noisy gradient in a probabilistic sense as $M\to\infty$
\begin{equation}
\begin{aligned}\tilde{\mathbf{p}}_{\text{u},i}-\mathbf{p}_{\text{u},i} & \triangleq\frac{\sum_{m=1}^{M}\tilde{G}_{i,m}^{2}\mathbf{c}_{m}}{\sum_{m=1}^{M}\tilde{G}_{i,m}^{2}}-\mathbf{p}_{\text{u},i}\\
 & =\frac{\sum_{m=1}^{M}\tilde{G}_{i,m}^{2}(\mathbf{c}_{m}-\mathbf{p}_{\text{u},i})}{\sum_{m=1}^{M}\tilde{G}_{i,m}^{2}}.
\end{aligned}
\label{eq:localization error}
\end{equation}
Recall that the noisy gradient is defined as 
\begin{equation}
\tilde{G}_{i,m}=\bar{G}_{i,m}+n_{i,m}\label{eq:noisy gradient-1}
\end{equation}
where $n_{i,m}\sim\mathcal{N}(0,\sigma_{i,m}^{2})$ is independent
Gaussian noise. Substituting (\ref{eq:noisy gradient-1}) into (\ref{eq:localization error}),
the localization error by adversary is then expanded to
\begin{equation}
\begin{aligned}\tilde{\mathbf{p}}_{\text{u},i}-\mathbf{p}_{\text{u},i} & =\frac{\sum_{m=1}^{M}(\bar{G}_{i,m}+n_{i,m})^{2}(\mathbf{c}_{m}-\mathbf{p}_{\text{u},i})}{\sum_{m=1}^{M}(\bar{G}_{i,m}+n_{i,m})^{2}}\\
 & =\frac{\sum_{m=1}^{M}(\bar{G}_{i,m}^{2}+2\bar{G}_{i,m}n_{i,m}+n_{i,m}^{2})(\mathbf{c}_{m}-\mathbf{p}_{\text{u},i})}{\sum_{m=1}^{M}(\bar{G}_{i,m}^{2}+2\bar{G}_{i,m}n_{i,m}+n_{i,m}^{2})}.
\end{aligned}
\label{eq:expandation}
\end{equation}
Here the squared noise term is decomposed as
\begin{equation}
\sum_{m=1}^{M}n_{i,m}^{2}=\sum_{m=1}^{M}\sigma_{i,m}^{2}+\sum_{m=1}^{M}(n_{i,m}^{2}-\sigma_{i,m}^{2}).\label{eq:noise}
\end{equation}
Thus, the complete expansion of the localization error is derived
as (\ref{eq:complete expansion}). 
\begin{figure*}
\begin{equation}
\tilde{\mathbf{p}}_{\text{u},i}-\mathbf{p}_{\text{u},i}=\frac{\sum_{m=1}^{M}(\bar{G}_{i,m}^{2}+\sigma_{i,m}^{2})(\mathbf{c}_{m}-\mathbf{p}_{\text{u},i})+2\sum_{m=1}^{M}\bar{G}_{i,m}n_{i,m}(\mathbf{c}_{m}-\mathbf{p}_{\text{u},i})+\sum_{m=1}^{M}(n_{i,m}^{2}-\sigma_{i,m}^{2})(\mathbf{c}_{m}-\mathbf{p}_{\text{u},i})}{\sum_{m=1}^{M}(\bar{G}_{i,m}^{2}+\sigma_{i,m}^{2})+2\sum_{m=1}^{M}\bar{G}_{i,m}n_{i,m}+\sum_{m=1}^{M}(n_{i,m}^{2}-\sigma_{i,m}^{2})}\label{eq:complete expansion}
\end{equation}
\end{figure*}
 In (\ref{eq:complete expansion}), $\sum_{m=1}^{M}(n_{i,m}^{2}-\sigma_{i,m}^{2})$
is the random fluctuation term with zero mean and variance
\begin{equation}
\mathrm{Var}\left(\sum_{m=1}^{M}(n_{i,m}^{2}-\sigma_{i,m}^{2})\right)=\sum_{m=1}^{M}\mathrm{Var}(n_{i,m}^{2})=\sum_{m=1}^{M}2\sigma_{i,m}^{4}.
\end{equation}
Similarly, the cross term $2\sum_{m=1}^{M}\bar{G}_{i,m}n_{i,m}$ has
zero mean and variance: 
\begin{equation}
\mathrm{Var}\left(2\sum_{m=1}^{M}\bar{G}_{i,m}n_{i,m}\right)=4\sum_{m=1}^{M}\bar{G}_{i,m}^{2}\sigma_{i,m}^{2}.
\end{equation}
Considering that the grid cells $\{\mathbf{c}_{m}\}$ are uniform
in a bounded region, with fixed total gradient power $\sum_{m=1}^{M}\bar{G}_{i,m}^{2}\sim O(1)$
and noise budget $\sum_{m=1}^{M}\sigma_{i,m}^{2}=\mu\sum_{m=1}^{M}\bar{G}_{i,m}^{2}\sim O(1)$,
the typical magnitudes are $\bar{G}_{i,m},\sigma_{i,m}\sim O(1/\sqrt{M})$.
Consequently, the variance of the cross term is
\begin{equation}
4\sum_{m=1}^{M}\bar{G}_{i,m}^{2}\sigma_{i,m}^{2}\sim O\left(4M\cdot\frac{1}{M^{2}}\right)=O\left(\frac{1}{M}\right).
\end{equation}
The variance of the quadratic fluctuation term is
\begin{equation}
\sum_{m=1}^{M}2\sigma_{i,m}^{4}\sim O\left(2M\cdot\frac{1}{M^{2}}\right)=O\left(\frac{1}{M}\right).
\end{equation}

By the Chebyshev's inequality, as $M\to\infty$, these fluctuation
terms $\sum_{m=1}^{M}(n_{i,m}^{2}-\sigma_{i,m}^{2})$ and $2\sum_{m=1}^{M}\bar{G}_{i,m}n_{i,m}$
vanish in probability. Thus, the attacker localization error is approximated
as
\begin{equation}
\tilde{\mathbf{p}}_{\text{u},i}-\mathbf{p}_{\text{u},i}\approx\frac{\sum_{m=1}^{M}(\bar{G}_{i,m}^{2}+\sigma_{i,m}^{2})(\mathbf{c}_{m}-\mathbf{p}_{\text{u},i})}{\sum_{m=1}^{M}(\bar{G}_{i,m}^{2}+\sigma_{i,m}^{2})}
\end{equation}
It is divided into two components as
\begin{equation}
\begin{aligned}\tilde{\mathbf{p}}_{\text{u},i}-\mathbf{p}_{\text{u},i}\quad\quad\quad\quad\quad\quad\quad\quad\quad\quad\quad\quad\quad\quad\quad\quad\quad\quad\quad\quad\\
\approx\frac{\sum_{m=1}^{M}\bar{G}_{i,m}^{2}(\mathbf{c}_{m}-\mathbf{p}_{\text{u},i})}{\sum_{m=1}^{M}(\bar{G}_{i,m}^{2}+\sigma_{i,m}^{2})}+\frac{\sum_{m=1}^{M}\sigma_{i,m}^{2}(\mathbf{c}_{m}-\mathbf{p}_{\text{u},i})}{\sum_{m=1}^{M}(\bar{G}_{i,m}^{2}+\sigma_{i,m}^{2})}
\end{aligned}
\end{equation}
With the maximum noise budget $\sum_{m=1}^{M}\sigma_{i,m}^{2}\leq\mu\sum_{m=1}^{M}\bar{G}_{i,m}^{2}$
in (\ref{eq:noise budget}), it is further reduced to
\begin{equation}
\begin{aligned}\tilde{\mathbf{p}}_{\text{u},i}-\mathbf{p}_{\text{u},i}\quad\quad\quad\quad\quad\quad\quad\quad\quad\quad\quad\quad\quad\quad\quad\quad\quad\quad\quad\quad\quad\\
\approx\frac{\sum_{m=1}^{M}\bar{G}_{i,m}^{2}(\mathbf{c}_{m}-\mathbf{p}_{\text{u},i})}{(1+\mu)\sum_{m=1}^{M}\bar{G}_{i,m}^{2}}+\frac{\sum_{m=1}^{M}\sigma_{i,m}^{2}(\mathbf{c}_{m}-\mathbf{p}_{\text{u},i})}{(1+\mu)\sum_{m=1}^{M}\bar{G}_{i,m}^{2}}\quad\\
=\frac{\sum_{m=1}^{M}\bar{G}_{i,m}^{2}(\mathbf{c}_{m}-\mathbf{p}_{\text{u},i})}{(1+\mu)\sum_{m=1}^{M}\bar{G}_{i,m}^{2}}+\frac{\mu\sum_{m=1}^{M}\sigma_{i,m}^{2}(\mathbf{c}_{m}-\mathbf{p}_{\text{u},i})}{(1+\mu)\sum_{m=1}^{M}\sigma_{i,m}^{2}}
\end{aligned}
\end{equation}
where
\begin{equation}
\Delta_{\text{g}}\triangleq\frac{\sum_{m=1}^{M}\bar{G}_{i,m}^{2}\mathbf{c}_{m}}{\sum_{m=1}^{M}\bar{G}_{i,m}^{2}}-\mathbf{p}_{\text{u},i}
\end{equation}
is the error component introduced by gradient clipping, and 
\begin{equation}
\Delta_{\text{n}}\triangleq\frac{\sum_{m=1}^{M}\sigma_{i,m}^{2}\mathbf{c}_{m}}{\sum_{m=1}^{M}\sigma_{i,m}^{2}}-\mathbf{p}_{\text{u},i}
\end{equation}
is the error component introduced by the added noise. Thus, the localization
error is simplified as
\begin{equation}
\tilde{\mathbf{p}}_{\text{u},i}-\mathbf{p}_{\text{u},i}\approx\frac{\Delta_{\text{g}}+\mu\Delta_{n}}{1+\mu}
\end{equation}
The proof is completed.

\bibliographystyle{IEEEtran}
\bibliography{JCgroup-bib/JCgroup-Jijia}

\end{document}